\documentclass[runningheads]{llncs}
\usepackage{graphicx}

\usepackage{tikz}
\usepackage{comment}
\usepackage{amsmath,amssymb} 
\usepackage{color}
\usepackage{multirow}
\usepackage{subfigure}

\usepackage[accsupp]{axessibility}  

\usepackage[width=122mm,left=12mm,paperwidth=146mm,height=193mm,top=12mm,paperheight=217mm]{geometry}

\begin{document}
\pagestyle{headings}
\mainmatter
\def\ECCVSubNumber{914}  

\title{An atrium segmentation network with location guidance and siamese adjustment} 


\titlerunning{An atrium segmentation network with location guidance and siamese adjustment}

\author{Yuhan Xie\inst{1} \and
Zhiyong Zhang\inst{1} \and
Shaolong Chen\inst{1} \and
Changzhen Qiu\inst{1,*}
}
\authorrunning{Xie et al.}
%
\institute{School of Electronics and Communication Engineering, Sun Yat-sen University, Guangzhou, China, qiuchzh@mail.sysu.edu.cn
}

\maketitle

\begin{abstract}
The segmentation of atrial scan images is of great significance for the three-dimensional reconstruction of the atrium and the
surgical positioning. Most of the existing segmentation networks adopt a 2D structure and only take original images as input, ignoring the context information of 3D images and the role of prior information. In this
paper, we propose an atrium segmentation network LGSANet with location guidance and siamese adjustment, which takes adjacent three slices of images as input and adopts an end-to-end approach to achieve
coarse-to-fine atrial segmentation. The location guidance(LG) block uses
the prior information of the localization map to guide the encoding features of the fine segmentation stage, and the siamese adjustment(SA)
block uses the context information to adjust the segmentation edges.
On the atrium datasets of ACDC and ASC, sufficient experiments prove
that our method can adapt to many classic 2D segmentation networks,
so that it can obtain significant performance improvements.

\end{abstract}

\section{Introduction}

Medical image segmentation of atrial region is of great significance for 3D reconstruction, pathological analysis, and surgical positioning based on atrium segmentation results. Before deep learning was widely used, many methods \cite{heimann2009statistical,engstrom2011segmentation,castro2015statistical,candemir2013lung,dodin2011fully}  based on traditional image processing were derived for segmentation. However, due to the noise in medical image imaging and the shape variability of organs in different cases, it is difficult for traditional methods to segment robustly and produce satisfactory results.

According to the understanding of the difficulty in atrium segmentation shown in \textbf{Figure 1}, we can find that the atrium images of MRI imaging have blurred boundaries (myocardium in the ACDC dataset \cite{acdc}) and large fluctuations in the boundary shape (left atrium in ASC dataset \cite{asc}). Therefore, how to use context information to assist in localization is a key point. After deep learning has been widely used, many networks with superior performance have emerged for medical image segmentation, such as: UNet \cite{ronneberger2015u}, UNet++ \cite{zhou2019unet++}, TransUNet \cite{chen2021transunet}, etc. From the perspective of contextual information utilization, most networks segment based on 2D slices, ignoring the contextual information in 3D images; while for 3D networks \cite{milletari2016v,zhou2021nnformer}, larger computing resources and larger datasets are often required, which is very difficult to achieve, so that it is difficult to use in some restricted scenarios.

\begin{figure}
	\centering
	\includegraphics[height=6.5cm]{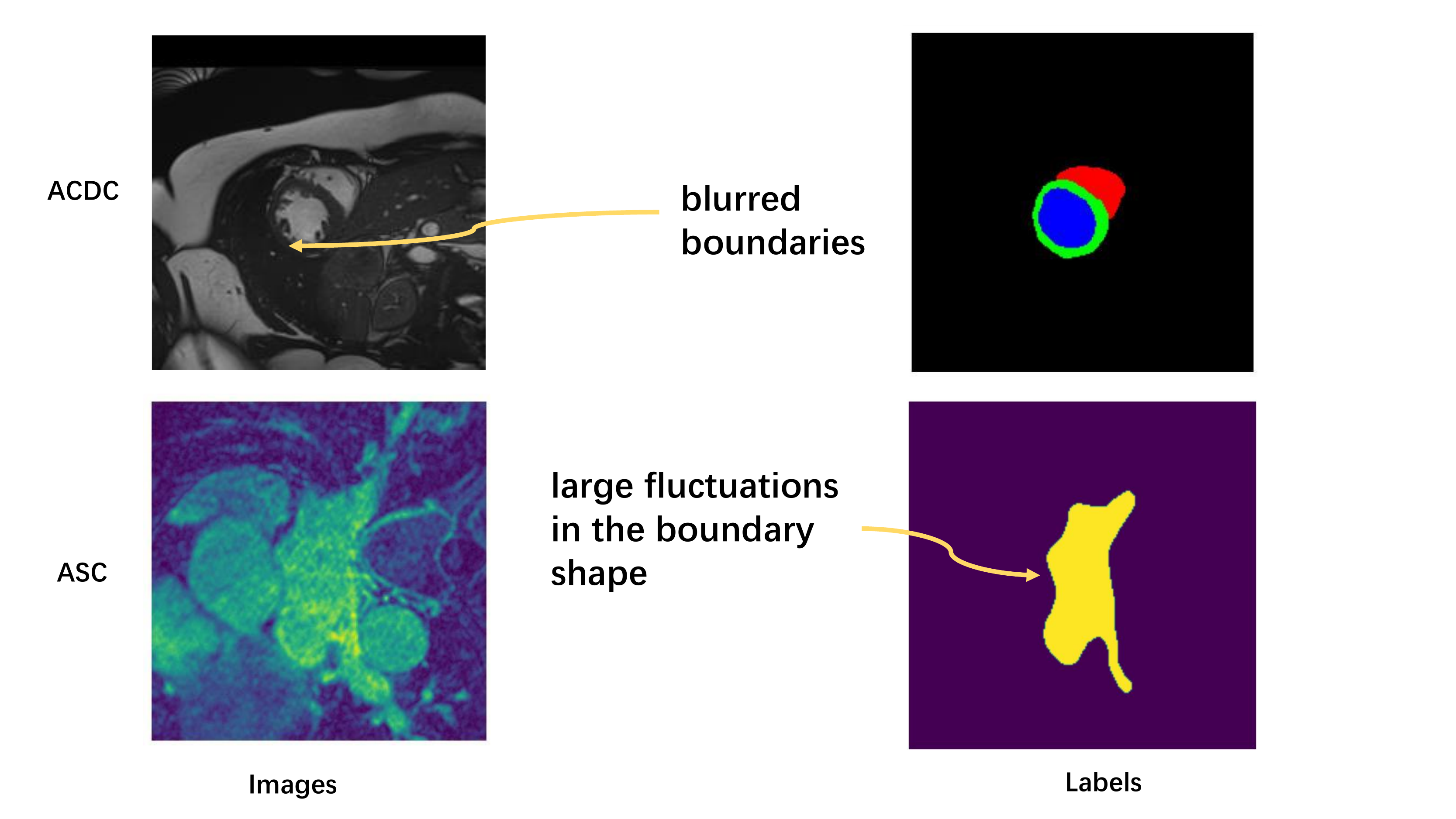}
	\caption{Diffiulty in segmenting atrial images.}
\end{figure}

From the perspective of physicians manually segmenting medical images, we often use vision for localization first, and then focus on the localization area for detailed segmentation. Some methods use a multi-stage method to obtain human like location and adjustment through tailoring between different stages. However, this method is not end-to-end, which brings difficulty in training and deployment; some methods simply use multiple networks to connect in series to achieve an end-to-end from coarse to fine segmentation, however, it does not take full advantage of coarse localization information.

Inspired by the above problems, we design an atrium segmentation network based on localization guidance and siamese adjustment: location guidance and siamese adjustment Network(LGSANet).

\textbf{Contributions:}

~

1. A two-stage end-to-end method is proposed, using location guidance(LG) block to utilize the coarse localization information and use it in the fine-tuning stage;

2. We adopt a siamese three-layer network structure for the segmentation of three-layer continuous slices, and use siamese adjustment(SA) block between the decoder layers to utilize context information, and fine-tune the segmentation edges through the continuity between slices;

3. The location guidance and siamese adjustment design can be fully applied to most existing excellent 2D networks to improve their performance,such as UNet and SwinUNet. Sufficient experiments have demonstrated the robustness and universality of our method.

\section{Related Work}
\subsection{Classic segmentation network}
The medical image segmentation methods can be mainly divided into the methods based on convolutional neural network \cite{ronneberger2015u,zhou2019unet++,huang2020unet,oktay2018attention,diakogiannis2020resunet,denseunet} and the methods based on transformer \cite{chen2021transunet,cao2021swin,zhou2021nnformer,hatamizadeh2022swin,xie2021cotr}. Among the methods based on convolutional neural network, UNet \cite{ronneberger2015u} in 2015 established the design direction of medical image segmentation network with the structure of classic encoder and decoder, and proved the effectiveness of skip connection; then UNet++ \cite{zhou2019unet++}, UNet3+ \cite{huang2020unet} explored different design of skip connection respectively to achieve a better interaction between encoder and decoder. While AttUNet \cite{oktay2018attention} introduces attention mechanism to the fusion of encoding and decoding features, ResUNet \cite{diakogiannis2020resunet} introduces residual design in convolution module, optimizING the design of UNet architecture from different directions. With the in-depth study of transformers \cite{vaswani2017attention}, the first medical segmentation network TransUNet \cite{chen2021transunet} that introduced transformers appeared, using the transformer architecture to realize the interaction of global information in deep semantic features. Then SwinUNet \cite{cao2021swin}, first medical segmentation network using pure transformers, proved the powerful representation capabilities of transformer.

\subsection{Coarse-to-fine segmentation}
The coarse-to-fine methods can be divided into mainly multi-stage methods \cite{astuto2021automatic} and end-to-end series connection methods \cite{ding2019stacked,hu2019s}. The former cuts the results of the first stage, and then performs the optimization in the second stage. This method is cumbersome ,complex and cannot provide an end-to-end solution; while the series connection method, like SMCSRNet \cite{ding2019stacked}, does not fully consider the positioning information in the coarse positioning stage, the simple connection may not lead to a good result.

\subsection{Network using context information}
MEPDNet \cite{mepdnet} uses a multi-encdoer and fusion decoder structure to utilize the context information, but it is directly fused in the decoder part which may lead to loss of context information; in addition,LSTM \cite{bai2018recurrent} is introduced to model the sequence relationship between the outputs of different slices, but too long-distance context information may increase the complexity of the model and the difficulty of training and deployment; ConResNet \cite{zhang2020inter} adopts a multi-task method, in which task 1 predicts the segmentation result, task 2 predicts the residual between slices.However,this method also requires great computing resources and large memory.

\section{Methods}

We characterize the 3D medical image as $M \in R^{C \times H \times W}$, take adjacent consecutive three-slice image as input, and describe it as $X=\left[S_{1}, S_{2}, S_{3}\right] \in R^{3 \times H \times W}$. Each slice is sent to the LGSA network in parallel, and the output $Y=\left[M_{1}, M_{2}, M_{3}\right] \in R^{3 \times H \times W}$ is obtained. The overall expression is shown in \textbf{Formula 1}:

\begin{align}
	Y=LGSA(X ; \theta)
\end{align}
Among them, we will select the output $M_{2}$ of the center slice as the final output.

\subsection{Overall structrue}

\begin{figure}
	\centering
	\includegraphics[height=7cm]{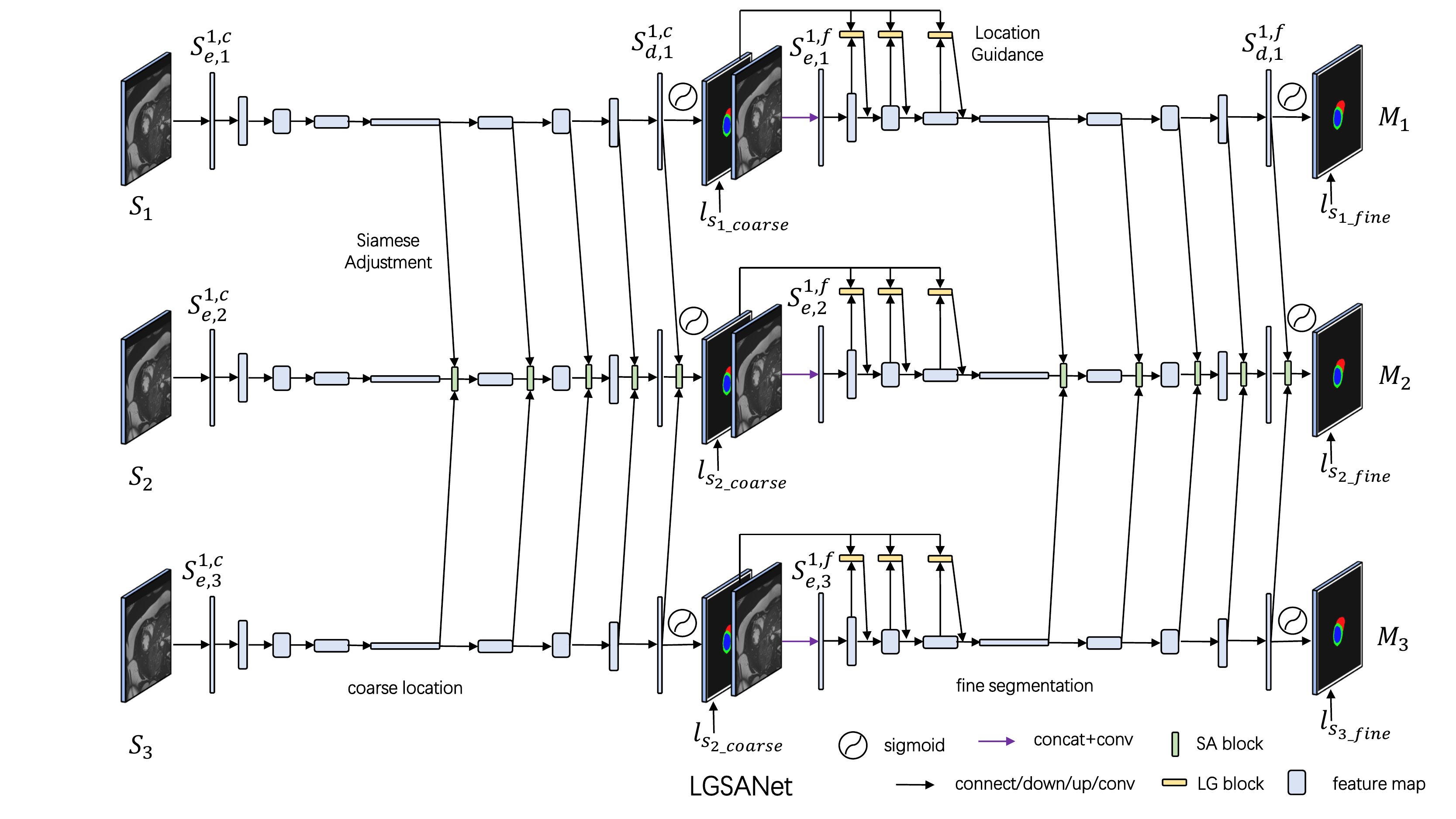}
	\caption{Overall Structure of LGSANet.}
\end{figure}

As shown in the \textbf{Figure 2}, we input consecutive three-layer 2D slices into three siamese single layer networks, each one is responsible for the segmentation of one-layer slice. The three single layer networks share parameters with each other to constrain the consistency of encoding and decoding. From the perspective of a single layer network, each one is divided into a coarse location and a fine segmentation stage. Between the two stages, location guidance(LG) blocks are used for fusion of multi-scale coarse location information and encoder features of fine segmentation stage, so that the model focuses on the located area in the fine segmentation stage. From the perspective of different slices, context information can be exchanged between different slices in the decoding stage. After each layer of decoding, an additional cross-slice siamese adjustment(SA) block will be performed so that the decoding features of the middle-layer can fully obtain context information, so as to use the continuity of the upper and lower slices for edge adjustment. The overall process of LGSANet can be expressed as \textbf{Formula 2}:

\begin{equation}
	\begin{gathered}
		S_{e, i}^{h, c}=C L_{\text {encoder }}^{h}\left(S_{e, i}^{h-1, c}\right), i=1,2,3 ; h=1,2 \ldots N \\
		S_{d, i}^{h-1, c}=C L_{\text {decoder }}^{h}\left(S_{a, i}^{h, c}\right), i=1,2,3 ; h=1,2 \ldots N \\
		S_{a, i}^{h, c}=C L_{S A}^{h}\left(S_{d, 1}^{h, c}, S_{d, 2}^{h, c}, S_{d, 3}^{h, c}\right), i=2 ; h=1,2 \ldots N \\
		S_{e, i}^{h, f}=F S_{\text {encoder }}^{h}\left(S_{l, i}^{h-1, f}\right), i=1,2,3 ; h=1,2 \ldots N \\
		S_{l, i}^{h, f}=F S_{L G}^{h}\left(S_{e, i}^{h, f}, L_{i}^{h}\right), i=1,2,3 ; h=1,2 \ldots N \\
		S_{d, i}^{h-1, f}=F S_{\text {decoder }}^{h}\left(S_{a, i}^{h, f}\right), i=1,2,3 ; h=1,2 \ldots N \\
		S_{a, i}^{h, f}=F G_{S A}^{h}\left(S_{d, i-1}^{h, f}, S_{d, i}^{h, f}, S_{d, i+1}^{h, f}\right), i=2 ; h=1,2 \ldots N
	\end{gathered}
\end{equation}
where $C L_{\text {encoder }}^{h}, C L_{\text {decoder }}^{h}, C L_{S A}^{h}, F G_{\text {encoder }}^{h}, F G_{L G}^{h}, F G_{\text {decoder }}^{h}, F G_{S A}^{h}$ represents different model components in LGSANet; $CL$ and $FS$ represent the coarse location and fine segmentation stages respectively; the superscript $h$ represents the h-th layer in the encoders or decoders, with a total of N layers, that is, N-1 times of downsampling are performed while N is 5 in UNet and 4 in SwinUNet. The subscripts $encoder$ and $decoder$ represent the encoder and decoder in this stage, while $SA$ and $LG$ represent the SA block and LG block in this stage. $ S_{e, i}^{h, c}, S_{d, i}^{h, c}, S_{e, i}^{h, f}, S_{d, i}^{h, f}, S_{a, i}^{h, c}, S_{a, i}^{h, f}$ represent the feature maps of different stages appearing in LGSANet; The superscript $c$ or $f$ indicates that the feature belongs to the coarse location or fine segmentation stage; the subscripts $e$, $f$, $a$, and $l$ indicate that they are the feature maps after the encoder, decoder, SA block, and LG block respectively;  the subscript $i$ indicates that it belongs to the feature map generated by the i-th slice.$L_{i}^{h}$ represents the multi-scale localization map generated in the coarse localization stage, the superscript $h$ represents the h-th layer, and the subscript $i$ represents the feature map generated by the i-th slice.

\subsection{Siamese feature encoding and decoding}
In the experiment, we mainly use UNet and SwinUNet as backbone. In the encoding process, UNet uses 5 layers of convolution modules as the basic units and performs 16 times downsampling, while SwinUNet uses 4 layers of swin transformer modules as the basic units, and completes 32 times downsampling through patch embedding and patch merging. In the decoding process , UNet uses a 4-layer convolution modules as the basic units and performs 16 times upsampling, while SwinUNet uses a 3-layer swin transformer modules as the basic units, and completes 32 times upsampling through patch expansion. For three-layer inputs, the encoder and decoder share parameters with each other. When performing siamese adjustment, SwinUNet needs to reshape features from vector form to feature map form. The main reasons why we use UNet and SwinUNet as backbones are: these two networks represent the most basic way of applying CNN and transformer in medical image segmentation, which composed of two different basic components; using them as backbones can effectively verify the performance robustness of our method in both cases.

\subsection{Location guidance block}

The location guidance block mainly uses the location information in stage 1 to guide the coding of the encoder features in stage 2 after multi-scale scaling. Through the introduction of prior information in localization map, the encoder features can be strengthened so that the model can pay more attention to the localization area. Its schematic diagram and formula are shown in \textbf{Formula 3,4} and \textbf{Figure 3}:

\begin{figure}
	\centering
	\includegraphics[height=7cm]{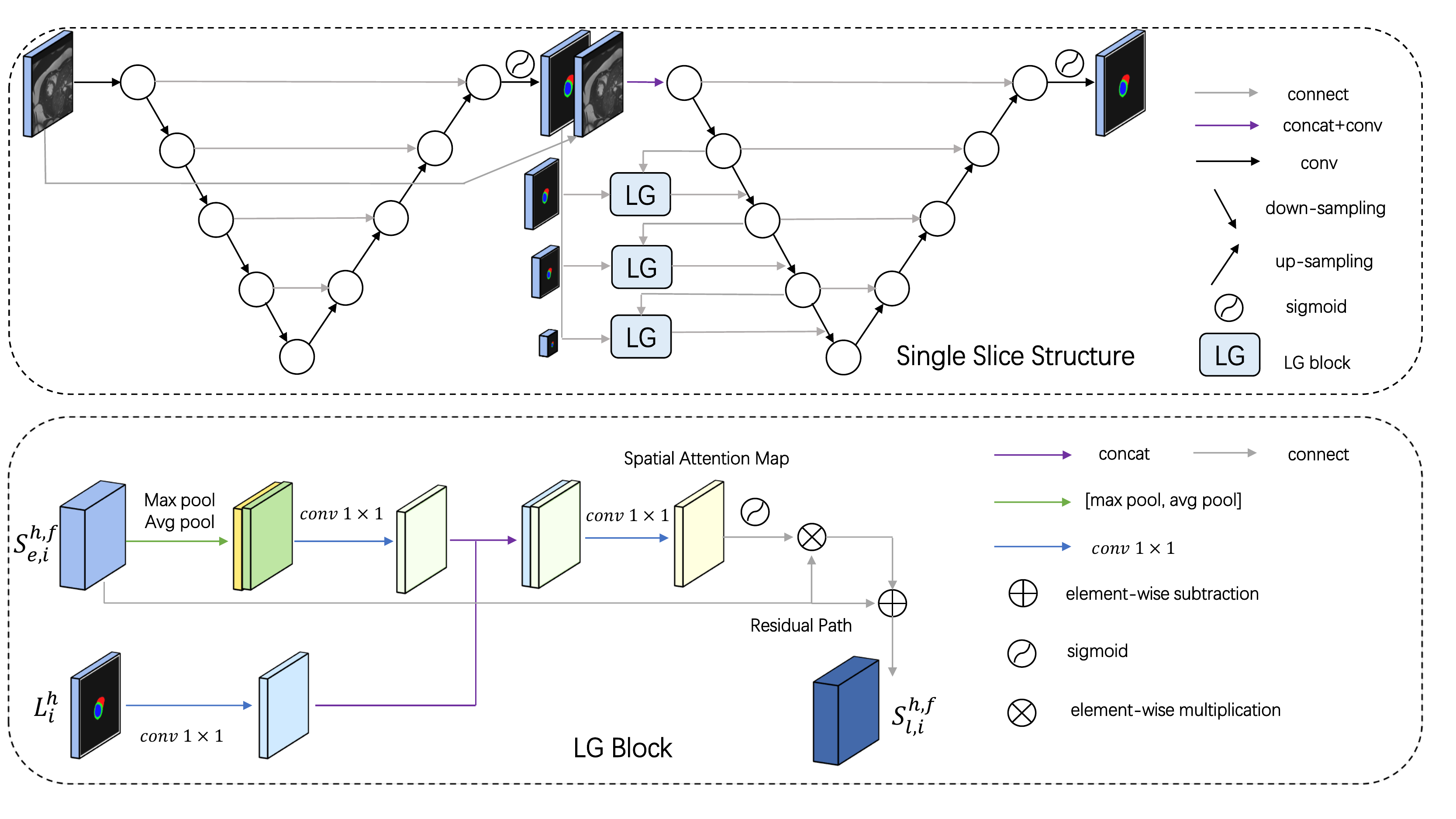}
	\caption{Location guidance block.}
\end{figure}

\begin{equation}
	{SA}_{map}={conv}_{1\times1}(concat(mp\left(S_{e,i}^{h,f}\right),ap\left(S_{e,i}^{h,f}\right),{conv}_{1\times1}(L_i^h)))
\end{equation}

\begin{equation}
	S_{l,i}^{h,f}=S_{e,i}^{h,f}\times softmax\left({SA}_{map}\right)+S_{e,i}^{h,f}
\end{equation}

where, $mp$ and $ap$ represent maxpooling and avgpooling respectively, and $S A_{m a p}$ represents the spatial attention map.

\subsection{Siamese adjustment block}
The siamese adjustment block mainly uses the context information of adjacent layers to adjust the output results of the intermediate layers. The input three-layer features are adjacently subtracted to obtain the edge differences, and adjacently multiplied to enlarge the overlapping areas. The edge differences are fused to obtain the edge feature and the overlapping areas are fused to obtain the central feature. Finally, the edge feature and central feature are fused together as output. The center branch uses the coincidence of the context to strengthen the center positioning of the middle layer, and the edge branch uses the edge continuity constraint of the context to fine-tune the edge of the middle layer. Its schematic diagram and formula are shown in  \textbf{Formula 5,6,7} and \textbf{Figure 4}:

\begin{figure}
	\centering
	\includegraphics[height=7cm]{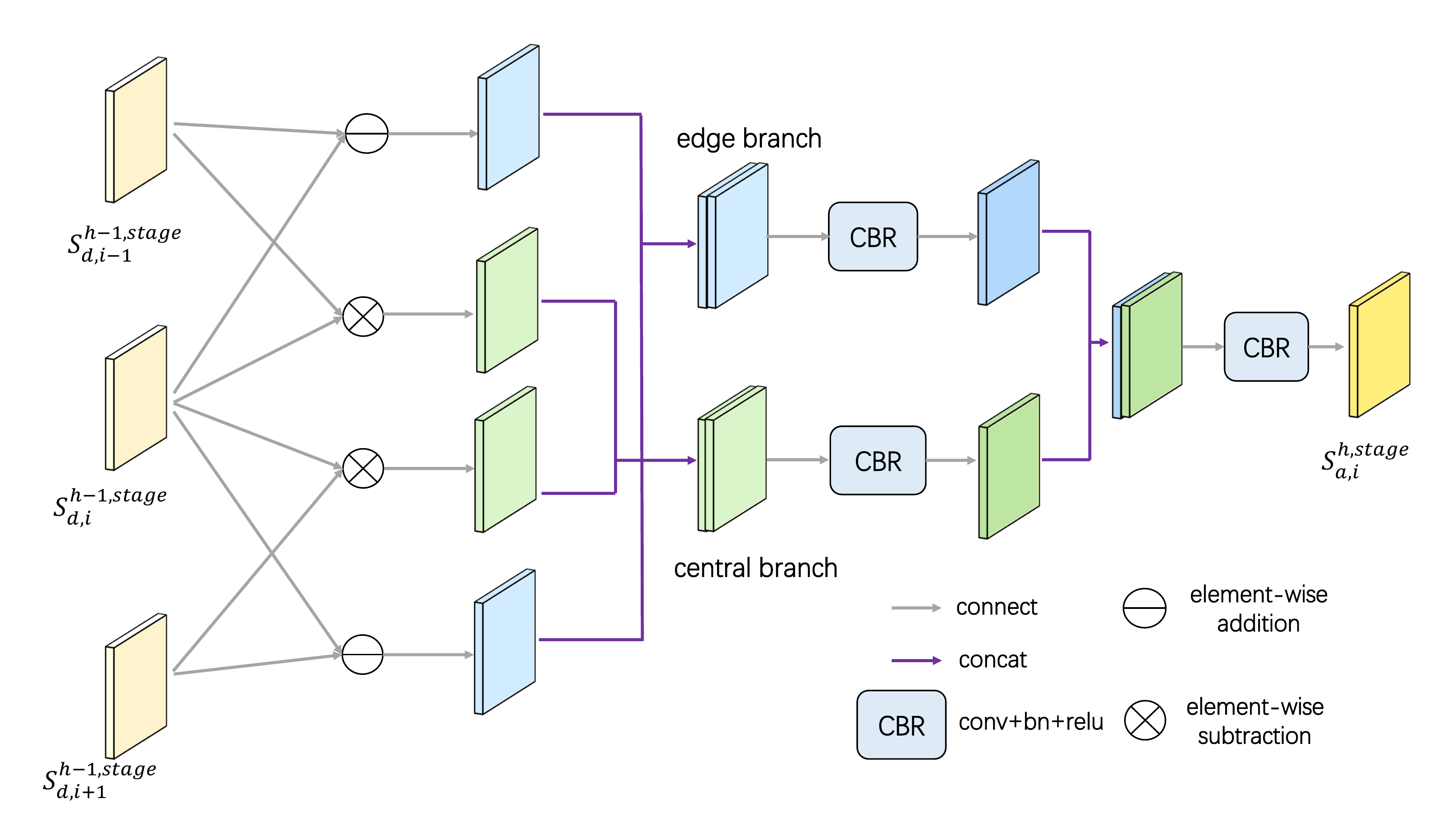}
	\caption{Siamese adjustment block.}
\end{figure}

\begin{equation}
	{S_{edge}=CBR(S}_{d,1}^{h,stage}-S_{d,2}^{h,stage},S_{d,2}^{h,stage}-S_{d,3}^{h,stage})
\end{equation}

\begin{equation}
	{S_{central}=CBR(S}_{d,1}^{h,stage}\times S_{d,2}^{h,stage},S_{d,2}^{h,stage}\times S_{d,3}^{h,stage})
\end{equation}

\begin{equation}
	S_{a,i}^{h,f}=CBR(S_{central},S_{edge})
\end{equation}

where, stage represents the coarse location or fine segmentation stage, and CBR represents the combination of conv3$\times$3, batch normalization, and Relu activation function.

\subsection{Serial supervision and siamese supervision}

In the design of the loss function, we use a combination of serial supervision and siamese supervision. While jointly supervising the coarse location and fine segmentation stages, the outputs of the three slices are all under siamese supervision to achieve the best optimization result. We use the output of the intermediate slices as the final output (except for the first and last layers) because it enjoys the most contextual information. The formula of the overall loss function can be expressed as formula 5,6,7:

\begin{equation}
	L_{all}={\beta L}_{coarse}+(1-\beta)L_{fine}
\end{equation}

\begin{equation}
	L_{coarse}={\alpha l}_{s_{1_coarse}}+(1-2\alpha)l_{s_{2_coarse}}+\alpha l_{s_{3_coarse}}
\end{equation}

\begin{equation}
	L_{fine}={\alpha l}_{s_{1_fine}}+(1-2\alpha)l_{s_{2_fine}}+\alpha l_{s_{3_fine}}
\end{equation}

\begin{equation}
	l_{s_{i_stage}}=\sum_{k\in S_i}\left(-\frac{1}{2}y_klog{\hat{y}}_k+1-\frac{2\times y_k\times{\hat{y}}_k}{y_k{+\hat{y}}_k}\right),stage=coarse,fine
\end{equation}

Where $k$ represents any point in slice i, and $y_{k}$, $\hat{y}_{k}$ represent the groundtruth and prediction result respectively. $\alpha$=0.33, $\beta$=0.5. This is because the quality of location and fine segmentation results, as well as the output results of different layers, affect each other. In order to ensure the final output of the midlle layer as good as possible, the location information needs to be accurate enough, and the adjacent layer outputs that provide fused interaction information also need to be reliable enough, so their weights are equally distributed. This will also be verified in the subsequent ablation experiments. The supervision of a single output is composed of dice loss and bce loss. The weight distribution of dice loss is higher than that of bce loss, because dice loss is more suitable for the segmentation of small targets, which can better overcome the imbalance of foreground and background.

\subsection{Structure comparison with different baselines}
\textbf{Figure 5} shows the comparison with baselines.In the design of network architecture, we mainly focus on the utilization of contextual information and coarse localization information. Therefore, from a coarse-to-fine point of view, the baseline we mainly refer to is SMCSRNet, which uses UNet with simple concatenation to complete end-to-end multi-stage segmentation. The difference is that we introduce a location guidance block in the middle of the two stages in order to make the fine-stage encoder pay more attention to the localization area. From the perspective of context information utilization, the baselines we mainly refer to are 3-slice UNet and MEPDNet. The 3-slice UNet uses continuous three-layer slices stacking as input,which is sent to a common encoder and decoder. MEPDNet uses three independent encoders extracting the features of the three-layer slices, and then performs fusion decoding. The difference is that we use siamese encoder and decoder to ensure the consistency and independence of encoding and decoding, and perform siamese adjustment between decoders to achieve information exchange at the same time. Subsequent experiments demonstrate that our design idea has gains in both aspects.
\begin{figure}
	\centering
	\includegraphics[height=7cm]{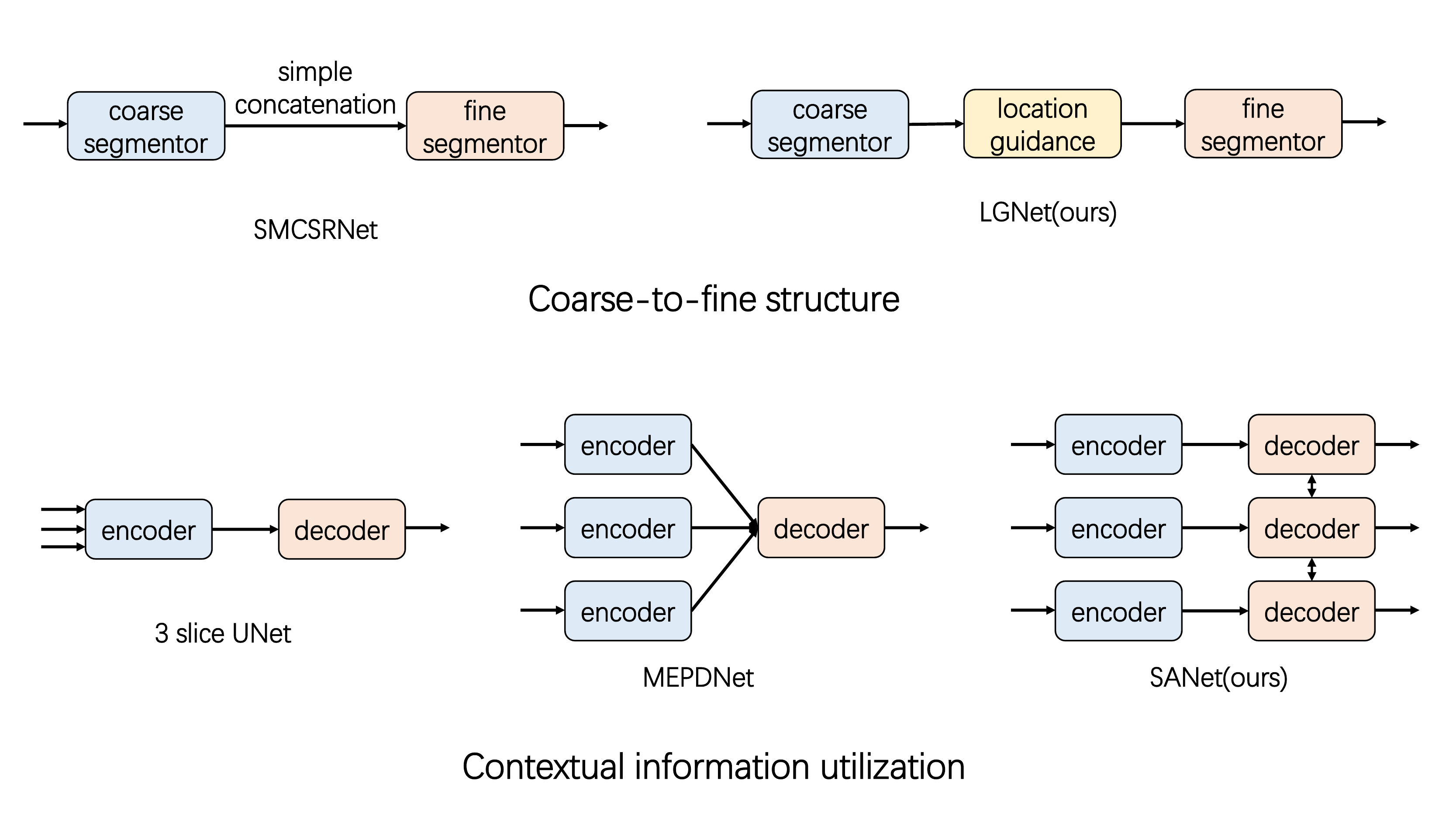}
	\caption{Structure comparison with baselines.}
\end{figure}

\section{Experimental results}
\subsection{Datasets}
Our framework mainly uses the ACDC dataset \cite{acdc}(2017 Automated cardiac diagnosis challenge) and ASC dataset \cite{asc} (the 2018 atrial segmentation challenge) for experiments. 

\textbf{ACDC}:The ACDC dataset contains 100 three-dimensional cardic MRI images to be segmented, each of which includes three types of manual annotations : left ventricle (LV), right ventricle (RV) and myocardium (MYO). Each case consists of a series of short-axis slices and the slice thickness is of 5 to 8 mm. The
short-axis in-plane spatial resolution goes from 0.83 to 1.75 $mm^{2}$/pixel.There are totally 951 slices included into experiments.

\textbf{ASC}:The ASC dataset contains 152 three dimensional MRI images for left atrium (LA) and each of which includes one type of annotation: left atrium(LA).The image resolution is 0.625 × 0.625 × 1.25 $mm^{3}$. There are totally 13552 slices included into experiments.

\subsection{Evaluation metrics}
We use the 95$\%$ Hausdorff Distance (HD95)(mm) , Dice score (DSC)($\%$), F1 score($\%$) to characterize the performance of the segmentation. The formula of DSC and F1 are shown in \textbf{Formula 12,13,14,15}:

\begin{equation}
	F1=2\times\frac{Precision\times R e c a l l}{Precision+Recall}
\end{equation}

\begin{equation}
	Recall=\frac{TP}{TP+FN}
\end{equation}

\begin{equation}
	Precision=\frac{TP}{TP+FP}
\end{equation}

\begin{equation}
	DSC=\ 2\times\frac{|X\cap Y|}{|X|+|Y|}
\end{equation}
where $X$ denotes the segmentation result of the method, and $Y$ denotes the ground truth.$TP$ denotes true negative result, $FP$ denotes false positive result and $FN$ denotes false negative result.

\subsection{Implementation details}
The training, validation and testing processes are all conducted on two RTX3090 cards.The approach is implemented by Python3.7 with Pytorch. the batch size is set to 24. An Adam optimizer is used in training process, with a learning rate of 5e-4,momentum of o.9 and weight decay of 1e-4. In the experiment,we train on ACDC for 100 epochs and 50 epochs on ASC, while the early stopping is set to be 20 epochs on ACDC and 10 epochs on ASC. Before conducting the experiments, we uniformly scale each slice to a size of 224$\times$224. The training set, validation set, and testing set are divided according to the ratio of 7:1:2. SwinUNet and its variants are initialized with pre-trained weights, and the rest of the models are initialized with Gaussian randomization. In order to ensure the reliability of the experimental results, we repeated the experiments for each category 5 times and obtained the average of the experimental results. We perform maximum and minimum normalization on the both datasets in preprocessing, which can be described by \textbf{Formula 16}:

\begin{equation}
	I(x,y)=\frac{I(x,y)-I_{min}}{I_{max}-I_{min}}
\end{equation}
where $I(x,y)$ denotes the grayscale of point (x,y), $Imax$ and $Imin$ denote the maximum value  and minimum value of an image.

In order to compare the difference between the output of the central layer of LGSANet and other methods, and to maintain the consistency of the comparison range, we let the head and tail slices of each 3D data not be included in the testing range; It is worth mentioning that our LGSANet can actually output the segmentation results of the first and last layers(this will be shown in our experimental results).

\subsection{Comparison with the state-of-the-art method}

We select CNN-based segmentation networks: UNet, UNet++, DenseUNet, ResUNet and transformer-based segmentation networks: SwinUNet, TransUNet for basic comparison. Besides, we also choose MEPDNet, SMCSRNet and 3-slice UNet as baselines from the aspects of contextual information ultilization and coarse-to-fine segmentation. UNet and SwinUNet are adopted as the two different backbones of our proposed LGSANet respectively. The experimental results are shown in \textbf{Table 1,2,3}:

\begin{table}[]
	\begin{center}
		\caption{Experiment results on ACDC dataset.}
		\label{table:headings}
		\begin{tabular}{ccc|ccc|ccc|ccc}
			\hline
			\multicolumn{3}{c|}{\multirow{2}{*}{Methods}}                                                                                                                                                                   & \multicolumn{3}{c|}{RV}                                                                                      & \multicolumn{3}{c|}{Myo}                                                                                     & \multicolumn{3}{c}{LV}                                              \\ \cline{4-12} 
			\multicolumn{3}{c|}{}                                                                                                                                                                                           & DSC                                & HD95                              & F1                                  & DSC                                & HD95                              & F1                                  & DSC                                & HD95          & F1             \\ \hline
			\multicolumn{1}{c|}{\multirow{9}{*}{\begin{tabular}[c]{@{}c@{}}One\\ slice\\ input\end{tabular}}}   & \multicolumn{1}{c|}{\multirow{6}{*}{\begin{tabular}[c]{@{}c@{}}One\\ stage\end{tabular}}} & ResUNet\cite{diakogiannis2020resunet}       & 88.45                              & 1.39                              & 87.18                               & 83.41                              & 1.37                              & 83.63                               & 91.37                              & 1.00          & 91.66          \\
			\multicolumn{1}{c|}{}                                                                               & \multicolumn{1}{c|}{}                                                                     & SwinUNet\cite{cao2021swin}      & 90.64                              & 1.29                              & 90.80                               & 83.82                              & 1.10                              & 84.09                               & 94.58                              & 0.67          & 94.68          \\
			\multicolumn{1}{c|}{}                                                                               & \multicolumn{1}{c|}{}                                                                     & UNet\cite{ronneberger2015u}          & 91.23                              & 1.24                              & 91.40                               & 84.56                              & 1.09                              & 84.77                               & 94.21                              & 0.46          & 94.35          \\
			\multicolumn{1}{c|}{}                                                                               & \multicolumn{1}{c|}{}                                                                     & UNet++\cite{zhou2019unet++}         & 91.58                              & 1.16                              & 91.69                               & 84.22                              & 1.14                              & 84.56                               & 94.48                              & 0.44          & 94.59          \\
			\multicolumn{1}{c|}{}                                                                               & \multicolumn{1}{c|}{}                                                                     & DenseUNet\cite{denseunet}     & 92.53                              & 1.06                              & 92.94                               & 84.89                              & 1.08                              & 84.96                               & 94.50                              & 0.39          & 94.84          \\
			\multicolumn{1}{c|}{}                                                                               & \multicolumn{1}{c|}{}                                                                     & TransUNet\cite{chen2021transunet}     & 92.28                              & 1.10                              & 92.59                               & 84.50                              & 1.11                              & 84.92                               & 95.39                              & 0.17          & 95.68          \\ \cline{2-12} 
			\multicolumn{1}{c|}{}                                                                               & \multicolumn{1}{c|}{\multirow{3}{*}{\begin{tabular}[c]{@{}c@{}}Two\\ stage\end{tabular}}} & SMCSRNet\cite{ding2019stacked}      & 91.99                              & 1.14                              & 92.12                               & 84.75                              & 1.08                              & 84.99                               & 95.14                              & 0.20          & 95.46          \\
			\multicolumn{1}{c|}{}                                                                               & \multicolumn{1}{c|}{}                                                                     & LG-SwinUNet   & 92.01                              & 1.12                              & 92.23                               & 84.89                              & 1.08                              & 85.12                               & 95.12                              & 0.21          & 95.34          \\
			\multicolumn{1}{c|}{}                                                                               & \multicolumn{1}{c|}{}                                                                     & LG-UNet       & \textbf{92.48}                     & \textbf{1.09}                     & \textbf{92.53}                      & \textbf{85.47}                     & \textbf{1.05}                     & \textbf{85.59}                      & \textbf{95.54}                     & \textbf{0.32} & \textbf{95.60} \\ \hline
			\multicolumn{1}{c|}{\multirow{6}{*}{\begin{tabular}[c]{@{}c@{}}Three\\ slice\\ input\end{tabular}}} & \multicolumn{1}{c|}{\multirow{4}{*}{\begin{tabular}[c]{@{}c@{}}One\\ stage\end{tabular}}} & 3-slice UNet  & 90.35          & 1.40          & 90.46          & 80.89          & 1.66         & 81.21         & 93.77          & 0.82          & 93.95          \\
			\multicolumn{1}{c|}{}                                                                               & \multicolumn{1}{c|}{}                                                                     & MEPDNet\cite{mepdnet}       & 91.53                              & 1.22                              & 91.52                               & 82.95                              & 1.23                              & 83.26                               & 94.31                              & 0.50          & 93.29          \\
			\multicolumn{1}{c|}{}                                                                               & \multicolumn{1}{c|}{}                                                                     & SA-SwinUNet   & 92.52          & 1.05          & 92.84         & 84.78         & 1.12          & 84.95         & 95.27         & 0.19          & 95.46          \\
			\multicolumn{1}{c|}{}                                                                               & \multicolumn{1}{c|}{}                                                                     & SA-UNet       & \textbf{92.94}                     & \textbf{1.00}                     & \textbf{93.11}                      & \textbf{86.65}                     & \textbf{1.00}                     & \textbf{86.84}                      & \textbf{95.53}                     & \textbf{0.17} & \textbf{95.70} \\ \cline{2-12} 
			\multicolumn{1}{c|}{}                                                                               & \multicolumn{1}{c|}{\multirow{2}{*}{\begin{tabular}[c]{@{}c@{}}Two\\ stage\end{tabular}}} & LGSA-SwinUNet & 92.92          & 1.02          & 93.05          & 85.51          & 1.05          & 85.08          & 95.73         & 0.11          & 95.90          \\
			\multicolumn{1}{c|}{}                                                                               & \multicolumn{1}{c|}{}                                                                     & LGSA-UNet     & \textbf{93.21} & \textbf{0.83} & \textbf{93.36} & \textbf{86.88} & \textbf{1.00}& \textbf{87.01} & \textbf{96.58} & \textbf{0.08} & \textbf{96.62} \\ \hline
		\end{tabular}
	\end{center}
\end{table}

\begin{table}[]
	\begin{center}
		\caption{Experiment results on ASC dataset.}
		\label{table:headings}
		\begin{tabular}{lll|lll}
			\hline
			\multicolumn{3}{l|}{Methods}                                                                                                                                                                                    & DSC            & HD95          & F1                                 \\ \hline
			\multicolumn{1}{l|}{\multirow{9}{*}{\begin{tabular}[c]{@{}l@{}}One\\ slice\\ input\end{tabular}}}   & \multicolumn{1}{l|}{\multirow{6}{*}{\begin{tabular}[c]{@{}l@{}}One\\ stage\end{tabular}}} & Resunet\cite{diakogiannis2020resunet}       & 82.00          & 2.52          & 83.01                              \\
			\multicolumn{1}{l|}{}                                                                               & \multicolumn{1}{l|}{}                                                                     & Swinunet\cite{cao2021swin}      & 87.53          & 1.70          & 88.17                              \\
			\multicolumn{1}{l|}{}                                                                               & \multicolumn{1}{l|}{}                                                                     & Unet\cite{ronneberger2015u}          & 90.53          & 1.26          & 90.79                              \\
			\multicolumn{1}{l|}{}                                                                               & \multicolumn{1}{l|}{}                                                                     & Unet++\cite{zhou2019unet++}        & 90.49          & 1.28          & 90.74                              \\
			\multicolumn{1}{l|}{}                                                                               & \multicolumn{1}{l|}{}                                                                     & DenseUNet\cite{denseunet}     & 89.92          & 1.32          & 90.42                              \\
			\multicolumn{1}{l|}{}                                                                               & \multicolumn{1}{l|}{}                                                                     & TransUNet\cite{chen2021transunet}     & 90.00          & 1.31          & 90.37                              \\ \cline{2-6} 
			\multicolumn{1}{l|}{}                                                                               & \multicolumn{1}{l|}{\multirow{3}{*}{\begin{tabular}[c]{@{}l@{}}Two\\ stage\end{tabular}}} & SMCSRNet\cite{ding2019stacked}      & 90.60          & 1.22          & 90.75                              \\
			\multicolumn{1}{l|}{}                                                                               & \multicolumn{1}{l|}{}                                                                     & LG-SwinUNet   & 89.12          & 1.52          & 89.45                              \\
			\multicolumn{1}{l|}{}                                                                               & \multicolumn{1}{l|}{}                                                                     & LG-UNet       & \textbf{90.81} & \textbf{1.21} & \multicolumn{1}{c}{\textbf{90.97}} \\ \hline
			\multicolumn{1}{l|}{\multirow{6}{*}{\begin{tabular}[c]{@{}l@{}}Three\\ slice\\ input\end{tabular}}} & \multicolumn{1}{l|}{\multirow{4}{*}{\begin{tabular}[c]{@{}l@{}}One\\ stage\end{tabular}}} & 3-slice UNet  & 90.12          & 1.32          & 90.34                              \\
			\multicolumn{1}{l|}{}                                                                               & \multicolumn{1}{l|}{}                                                                     & MEPDNet\cite{mepdnet}       & 90.18          & 1.21          & 90.66                              \\
			\multicolumn{1}{l|}{}                                                                               & \multicolumn{1}{l|}{}                                                                     & SA-SwinUNet   & 89.82          & 1.36          & 89.98                              \\
			\multicolumn{1}{l|}{}                                                                               & \multicolumn{1}{l|}{}                                                                     & SA-UNet       & \textbf{91.57} & \textbf{1.09} & \textbf{91.74}                     \\ \cline{2-6} 
			\multicolumn{1}{l|}{}                                                                               & \multicolumn{1}{l|}{\multirow{2}{*}{\begin{tabular}[c]{@{}l@{}}Two\\ stage\end{tabular}}} & LGSA-Swinunet & 90.17          & 1.30          & 90.51                              \\
			\multicolumn{1}{l|}{}                                                                               & \multicolumn{1}{l|}{}                                                                     & LGSA-unet     & \textbf{91.85} & \textbf{1.02} & \textbf{92.06}                     \\ \hline
		\end{tabular}
	\end{center}
\end{table}

As shown in \textbf{Table 1} and \textbf{Table 2}, LGSAUNet achieved the best segmentation results in both ACDC and ASC datasets, and performed the best in DSC, HD95, and F1 indicators. It can be seen that the segmentation effect has been significantly improved by using LGSANet structure. On the three targets of RV, Myo, and LV in ACDC, LGSAUNet obtained $1.98\% $, $2.32\%$, and $2.37\%$ dice performance improvement compared with UNet respectively. On the ASC dataset, LGSAUNet obtained 1.32$\%$ performance improvement compared to UNet. As a variant of LGSANet, LGSA-Swinunet has also improved significantly, with 2.28$\%$, 1.69$\%$, 1.15$\%$ on ACDC, and 2.64$\%$ on ASC. It can be seen that the design structure of LGSANet has good applicability to the architecture of both CNN and transformer.

For the models that simply using one slice as input and adopt two stage optimization, LGNet that using location guidance achieve better improvements compared with the SMCSRNet that using simply UNet concatenation.For the models that using three slices as input but simply using one stage optimzation,SANet that using siamese adjustment also performs better than 3-slice UNet and MEPDNet.

\begin{table}[]
	\begin{center}
		\caption{The dice($\%$) of different output in LGSANet.}
		\label{table:headings}
		\begin{tabular}{cc|ccc|ccc}
			\hline
			\multicolumn{2}{c|}{\multirow{2}{*}{Datasets}} & \multicolumn{3}{c|}{Coarse location}                    & \multicolumn{3}{c}{Fine segmentation}                  \\ \cline{3-8} 
			\multicolumn{2}{c|}{}                          & \multicolumn{1}{c}{1} & \multicolumn{1}{c}{2} & 3     & \multicolumn{1}{c}{1} & \multicolumn{1}{c}{2} & 3     \\ \hline
			\multicolumn{1}{c|}{ACDC}         & RV         & 91.34                  & 92.89                  & 90.72 & 91.71                  & \textbf{93.21}                  & 91.34 \\ \cline{2-2}
			\multicolumn{1}{c|}{}             & Myo        & 86.23                  & 86.60                  & 84.55 & 86.62                  & \textbf{86.88}                  & 84.89 \\ \cline{2-2}
			\multicolumn{1}{c|}{}             & LV         & 95.75                  & 96.19                  & 95.03 & 96.08                  & \textbf{96.58}                  & 95.48 \\ \cline{1-2}
			\multicolumn{2}{c|}{ASC}                       & 90.89                  & 91.47                  & 90.77 & 91.31                  & \textbf{91.85}                  & 91.26 \\ \hline
		\end{tabular}
	\end{center}
\end{table}

From the results in \textbf{Table 3}, it can be seen that the output results of the middle layer slices are better than the adjacent layers.The output results of the fine segmentation stage are better than the results of the coarse location stage as well. It illustrates that the idea of optimizing the central layer from coarse to fine with the help of context information actually works. Besides,for the output of the adjacent layers in fine segmentation, although it is a little worse than the center layer, is still better than the output of its 2D basic network, so it can also be benefit for the segmentation of the first and last slices.The mean and fluctuation range of the experimental results are shown in \textbf{Figure 6}.

\begin{figure}[h]
	\centering
	\begin{minipage}{0.45\linewidth}
		\vspace{3pt}
		\centerline{\includegraphics[width=\textwidth]{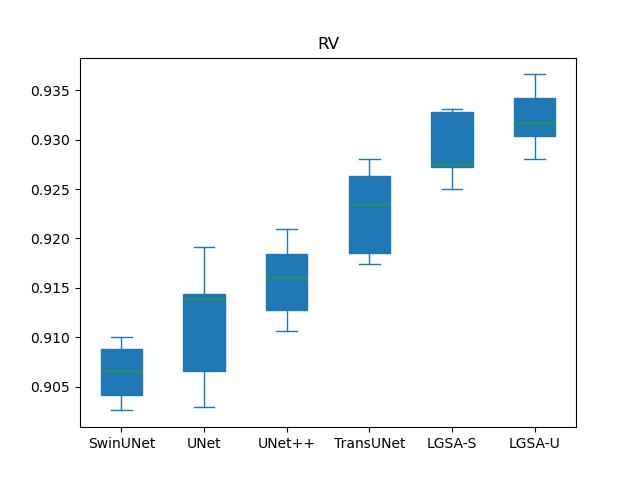}}
		\vspace{3pt}
		\centerline{\includegraphics[width=\textwidth]{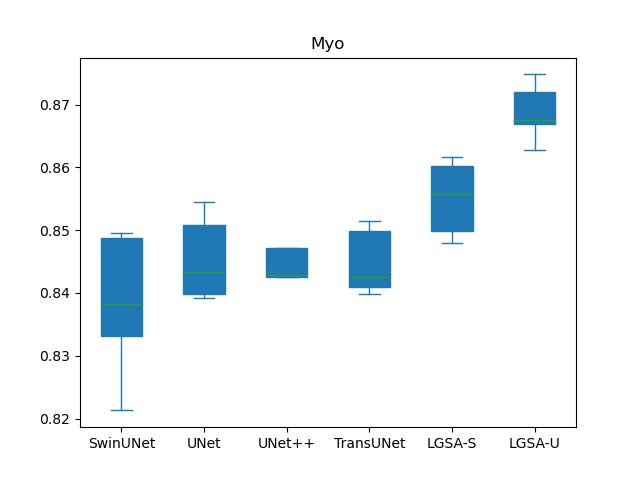}}
	\end{minipage}
	\begin{minipage}{0.45\linewidth}
		\vspace{3pt}
		\centerline{\includegraphics[width=\textwidth]{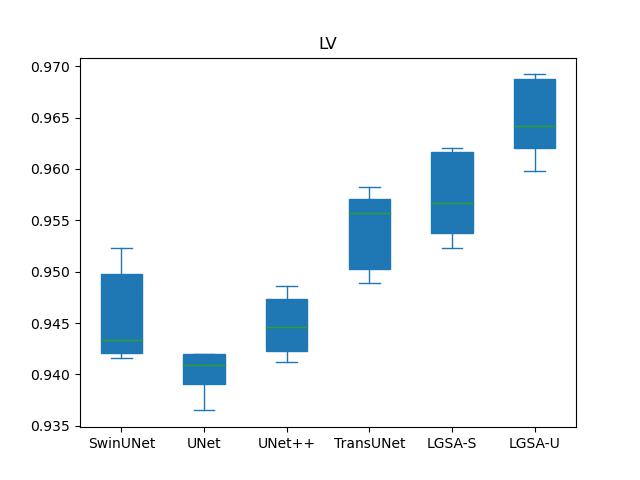}}
		\vspace{3pt}
		\centerline{\includegraphics[width=\textwidth]{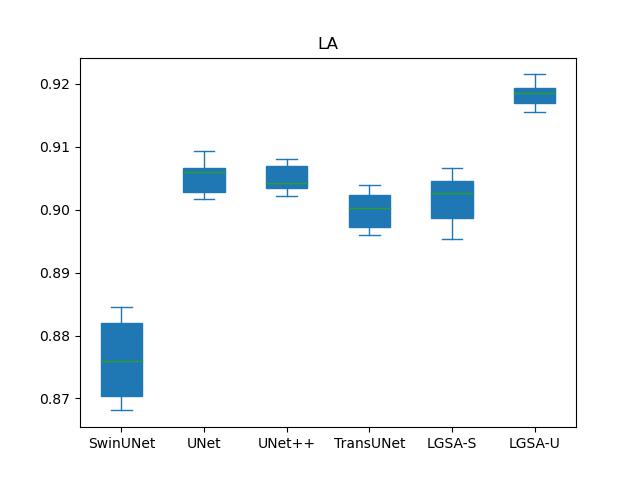}}
	\end{minipage}
	\caption{The box and whisker plot on ACDC(LV,Myo,RV) and ASC(LA) dataset.}
\end{figure}

\subsection{Ablation Analysis of Our Method}

\begin{table}[]
	\begin{center}
		\caption{Ablation on different modules used in LGSANet.}
		\label{table:headings}
		\begin{tabular}{c|ccc|ccc}
			\hline
			\multirow{2}{*}{Methods} & \multicolumn{3}{c|}{ACDC}                                    & \multicolumn{3}{c}{ASC}                                      \\ \cline{2-7} 
			& \multicolumn{1}{c}{DSC} & \multicolumn{1}{c}{HD95} & F1    & \multicolumn{1}{c}{DSC} & \multicolumn{1}{c}{HD95} & F1    \\ \hline
			UNet                     & 90.00                    & 0.93                      & 90.17 & 90.53                    & 1.26                      & 90.79 \\
			UNet+LG                  & 91.16                    & 0.82                      & 91.24 & 90.81                    & 1.21                      & 90.97 \\
			UNet+SA                  & 91.70                    & 0.72                      & 91.88 & 91.57                    & 1.09                      & 91.74 \\
			UNet+LG+SA               & \textbf{92.22}                    & \textbf{0.64}                      & \textbf{92.33} & \textbf{91.85}                    & \textbf{1.02}                      & \textbf{92.06} \\ \hline
		\end{tabular}
	\end{center}
\end{table}

\begin{table}[]	
	\begin{center}
		\caption{Ablation on different design of SA block.}
		\label{table:headings}
		\begin{tabular}{c|ccc}
			\hline
			\multirow{2}{*}{Methods} & \multicolumn{3}{c}{ACDC}                                     \\ \cline{2-4} 
			& \multicolumn{1}{c}{DSC} & \multicolumn{1}{c}{HD95} & F1    \\ \hline
			Multi-head               & 91.90                    & 0.68                      & 91.99 \\
			Central-head             & \textbf{92.22}                    & \textbf{0.64}                      & \textbf{92.33} \\ \hline
		\end{tabular}
	\end{center}
\end{table}

\begin{figure}
	\centering
	\includegraphics[height=7cm]{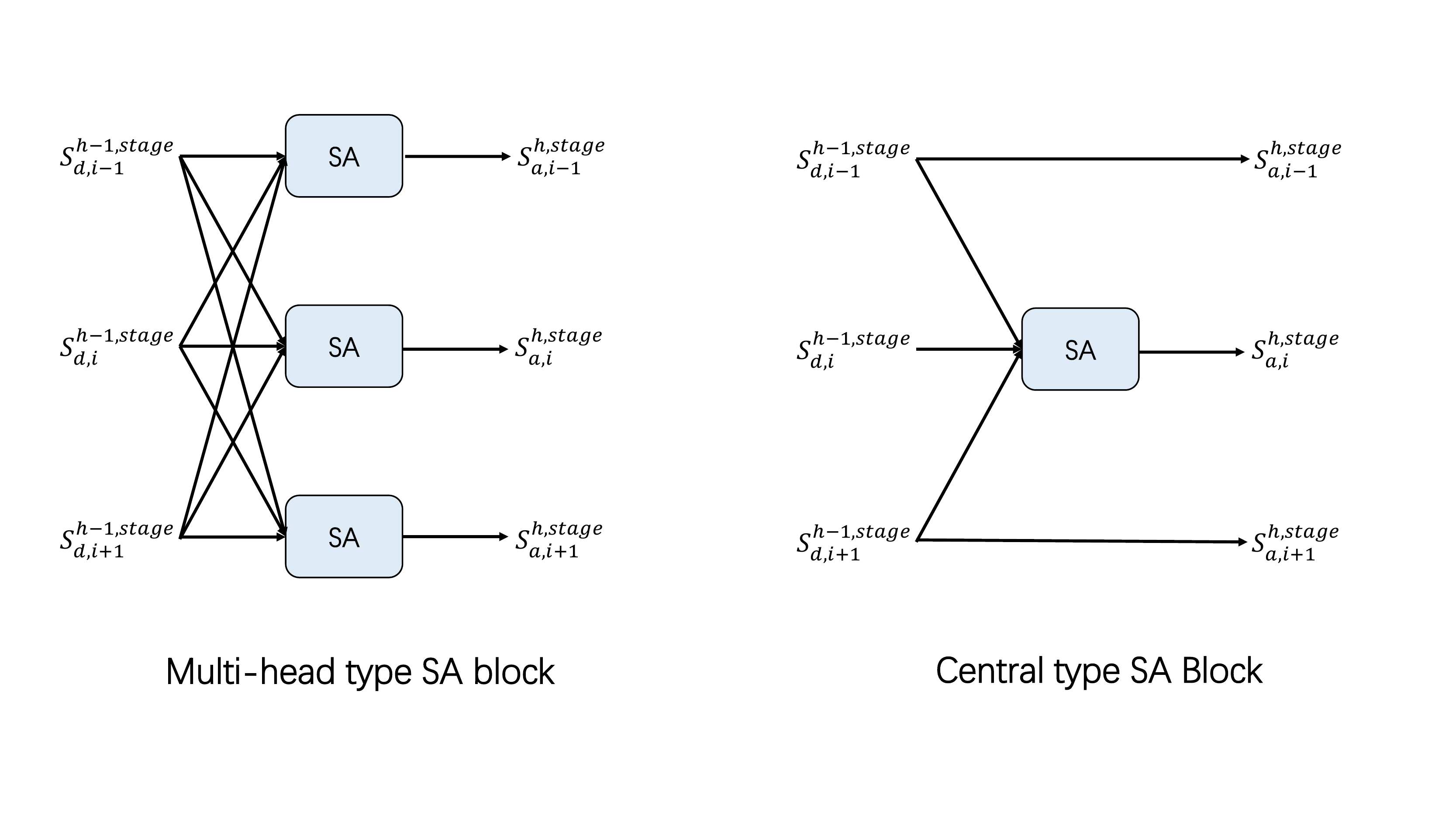}
	\caption{Multi-head type and central type SA block.}
\end{figure}

From the results in \textbf{Table 4}, it can be seen that the use of LG block and SA block can gradually improve the performance of the segmentation network. As shown in \textbf{Figure 6}, it can be seen that the single-head output results can obtain better benefits than the multi-head output.In multi-head mode, it is probably unreasonable to use the edge information to correct the edge layer features.But the middle layer in the single-head mode can obtain balanced context information, so the effect is relatively better.As shown in \textbf{Table 5}, the increase in the number of SI blocks enables both the deep and shallow contextual information to be acquired and adjusted during the decoding process, which also shows that the SA block and skip connection in the 2d network play a similar role. As shown in \textbf{Table 7}, the combination of serial supervision and siamese supervision enables LGSANet to gradually obtain better performance under the structure of LGSANet.

\begin{table}[]
	\centering
	\caption{Ablation on the number of SA block. The number of SA block veries from 1 to 5 in LGSA-UNet.}
	\label{table:headings}
	\begin{tabular}{c|ccc}
		\hline
		\multirow{2}{*}{\begin{tabular}[c]{@{}c@{}}Number of\\ SA block\end{tabular}} & \multicolumn{3}{c}{ACDC}                                     \\ \cline{2-4} 
		& \multicolumn{1}{c}{DSC} & \multicolumn{1}{c}{HD95} & F1    \\ \hline
		1                                                                             & 91.27                    & 0.78                      & 91.54 \\
		3                                                                             & 91.92                    & 0.68                      & 92.05 \\
		5                                                                             & \textbf{92.22}                    & \textbf{0.64}                      & \textbf{92.33} \\ \hline
	\end{tabular}
\end{table}	

\begin{table}[]
	\centering
	\caption{Ablation on loss function. OS repesents that only ouput of central slice in fine segmentation stage is supervised; SiS means serial supervision and Sis means siamese supervision. }
	\label{table:headings}
	\begin{tabular}{c|ccc}
		\hline
		\multirow{2}{*}{\begin{tabular}[c]{@{}c@{}}Supervision\\ type\end{tabular}} & \multicolumn{3}{c}{ACDC}                                     \\ \cline{2-4} 
		& \multicolumn{1}{c}{DSC} & \multicolumn{1}{c}{HD95} & F1    \\ \hline
		OS                                                                          & 90.57                    & 0.88                      & 90.84 \\
		SeS                                                                          & 90.89                    & 0.79                      & 91.21 \\
		SiS                                                                          & 91.84                    & 0.70                      & 91.97 \\
		SeS+SiS                                                                       & \textbf{92.22}                    & \textbf{0.64}                      & \textbf{92.33} \\ \hline
	\end{tabular}
\end{table}

\section{Visualization}

It can be seen from the visualization results that our approach LGSA-UNet reach the best proformance.In the two datasets, our method can accurately segment the target, making the segmentation result smoother and more accurate. In ACDC dataset, the discontinuity of the segmentation edge is greatly reduced. ASC is a dataset with rich target morphological changes, our method can also better fit the boundaries of the target. Compared with its 2D backbone model, LGSANet can achieve great performance improvement in segmentation task.

\begin{figure}[h]
	\centering
	\begin{minipage}{0.15\linewidth}
		\vspace{3pt}
		\centerline{\includegraphics[width=\textwidth]{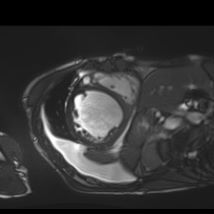}}
		\vspace{3pt}
		\centerline{\includegraphics[width=\textwidth]{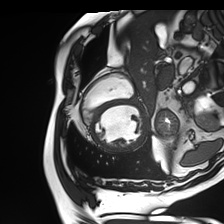}}
		\vspace{3pt}
		\centerline{\includegraphics[width=\textwidth]{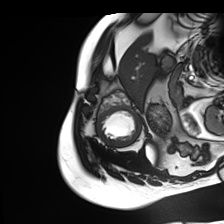}}
		\centerline{Image}
	\end{minipage}
	\begin{minipage}{0.15\linewidth}
		\vspace{3pt}
		\centerline{\includegraphics[width=\textwidth]{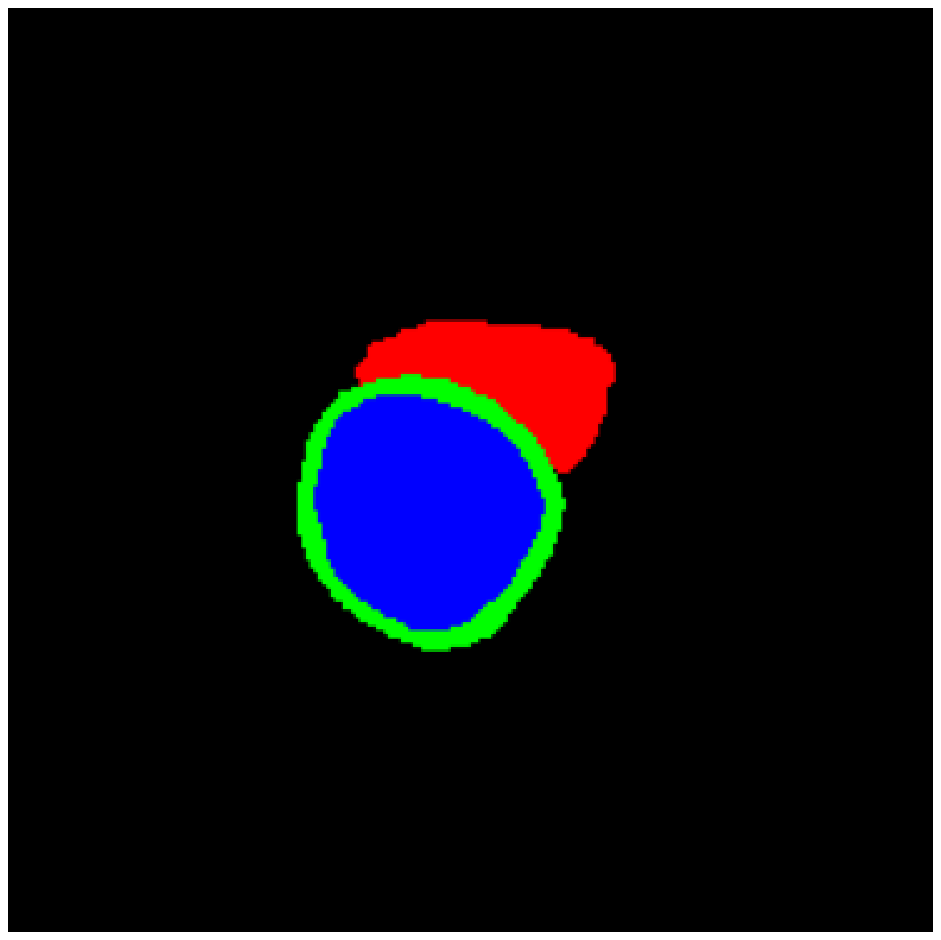}}
		\vspace{3pt}
		\centerline{\includegraphics[width=\textwidth]{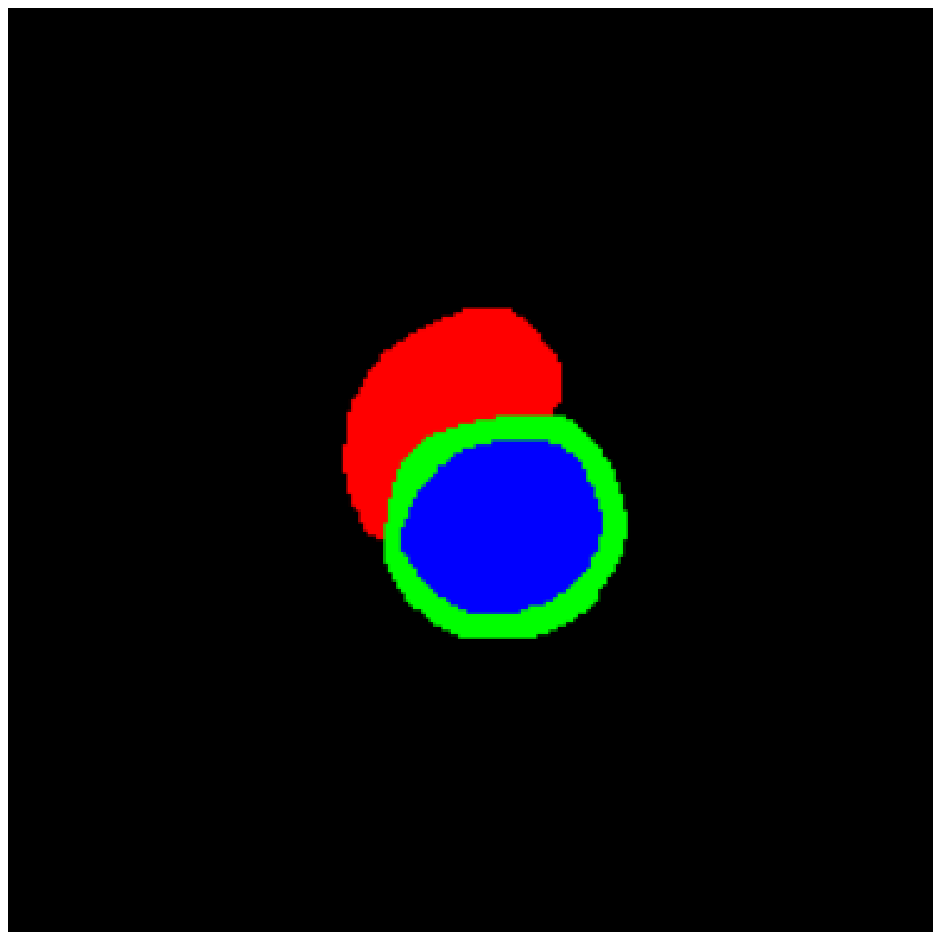}}
		\vspace{3pt}
		\centerline{\includegraphics[width=\textwidth]{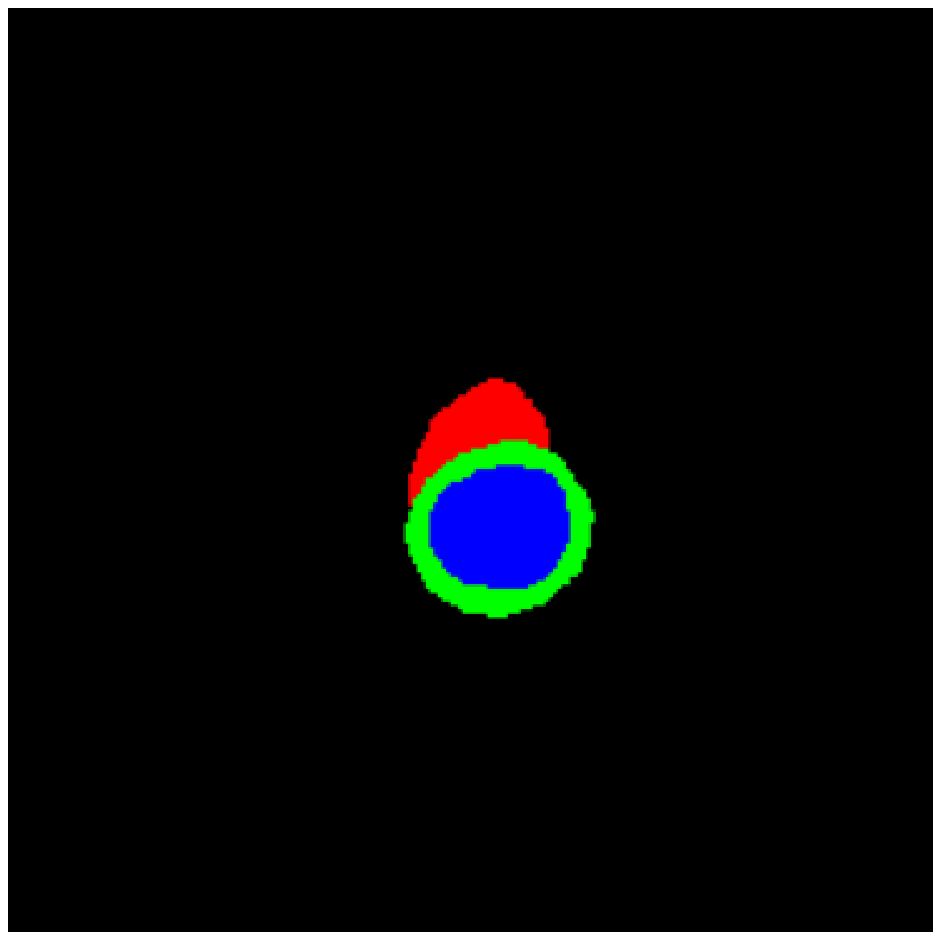}}
		\centerline{GroundTruth}
	\end{minipage}
	\begin{minipage}{0.15\linewidth}
		\vspace{3pt}
		\centerline{\includegraphics[width=\textwidth]{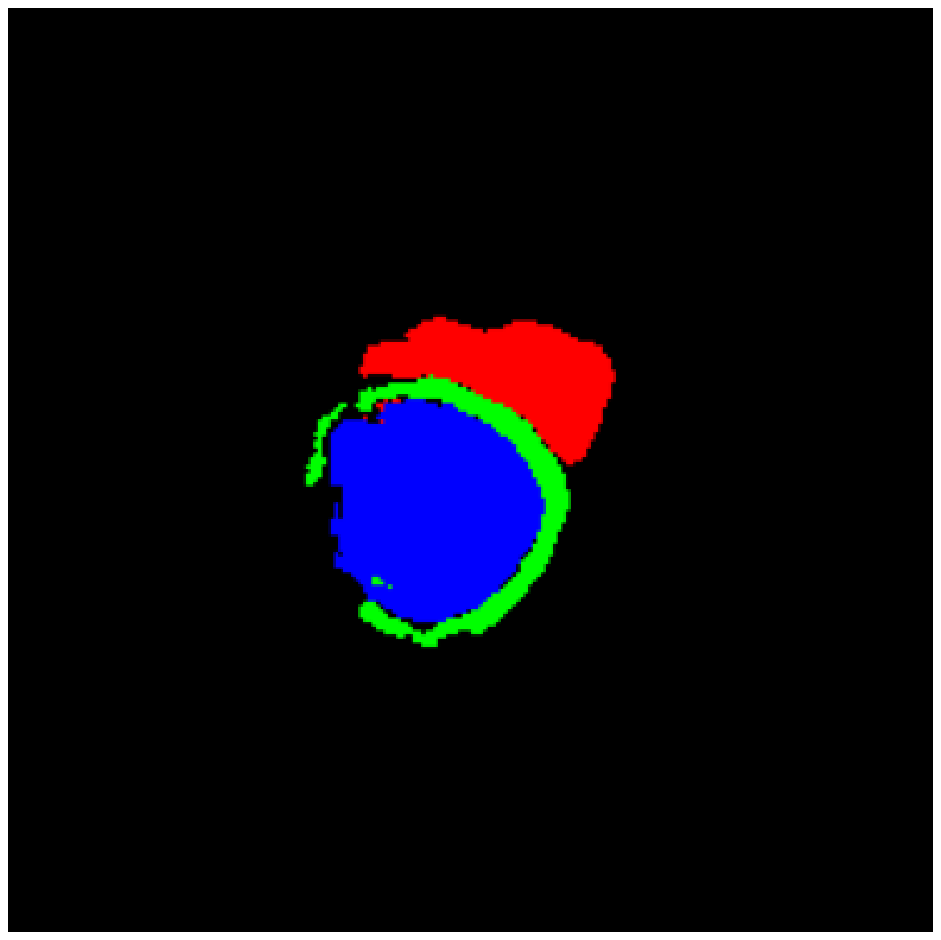}}
		\vspace{3pt}
		\centerline{\includegraphics[width=\textwidth]{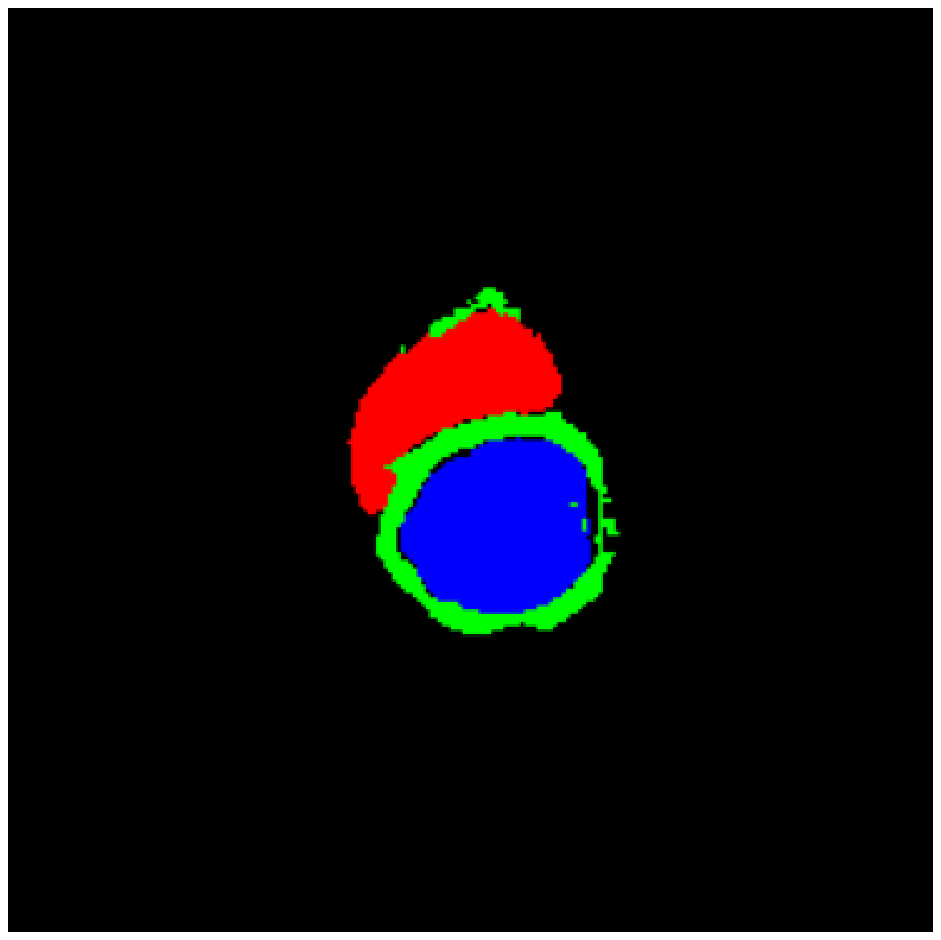}}
		\vspace{3pt}
		\centerline{\includegraphics[width=\textwidth]{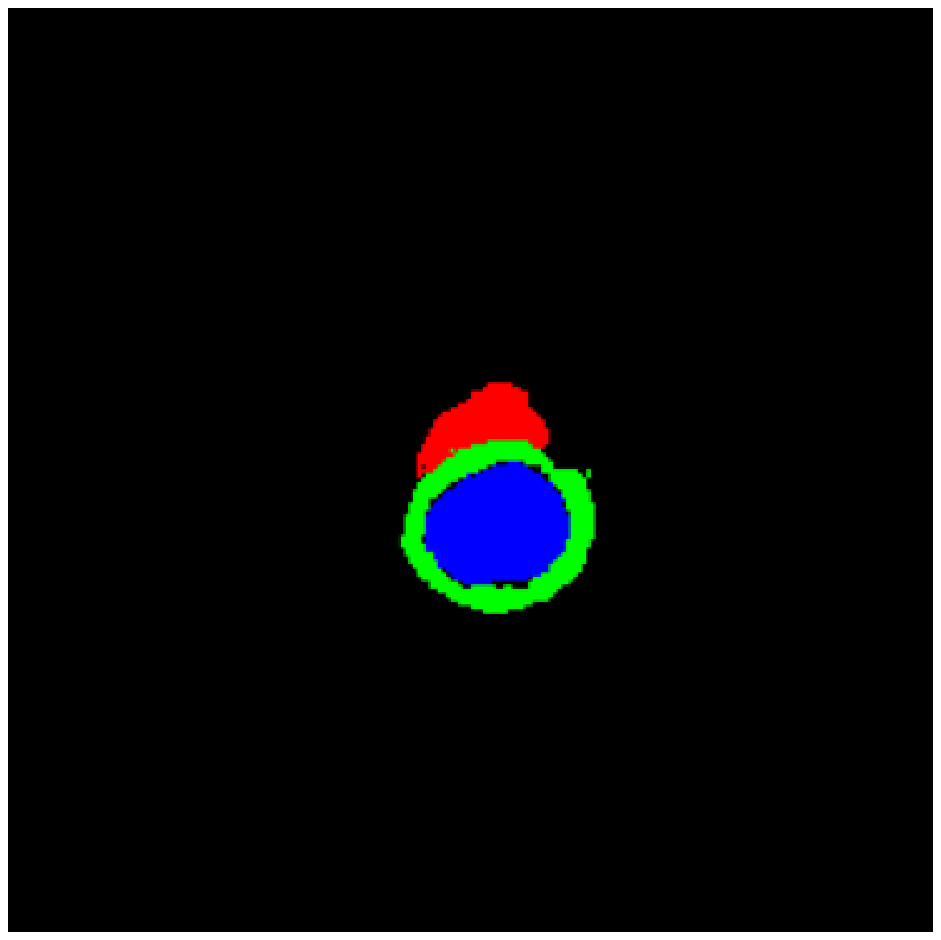}}
		\centerline{LGSA-S}
	\end{minipage}
	\begin{minipage}{0.15\linewidth}
		\vspace{3pt}
		\centerline{\includegraphics[width=\textwidth]{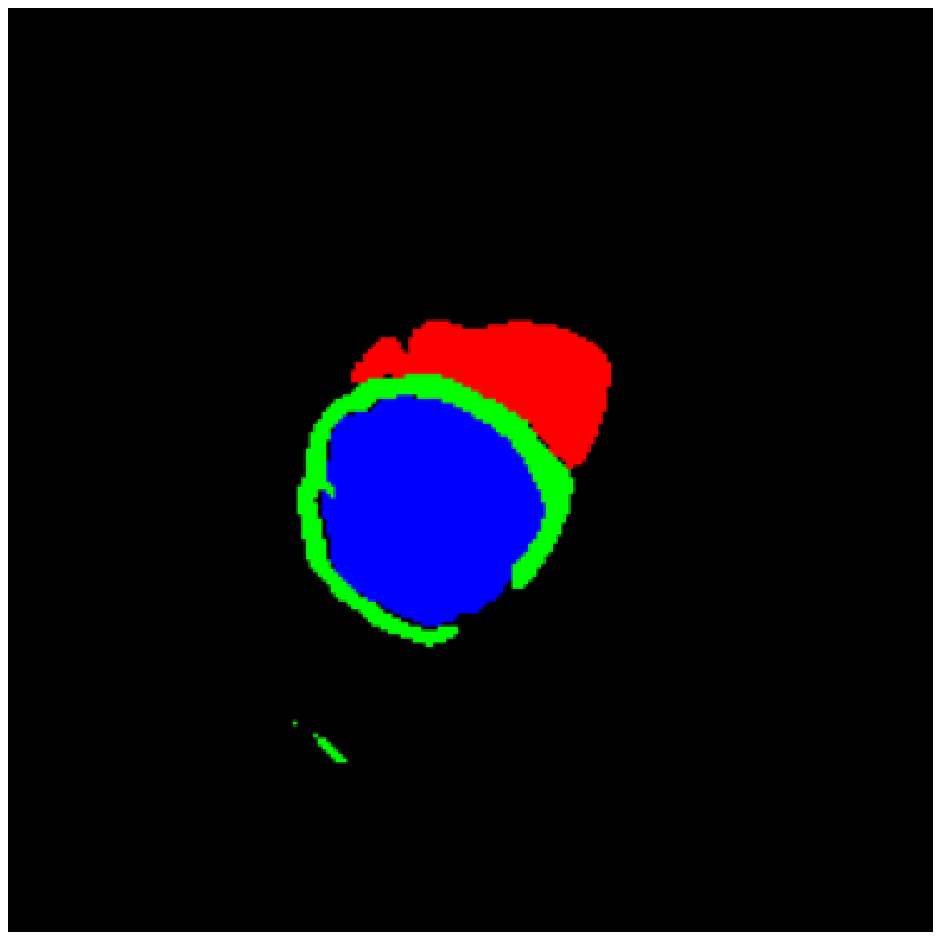}}
		\vspace{3pt}
		\centerline{\includegraphics[width=\textwidth]{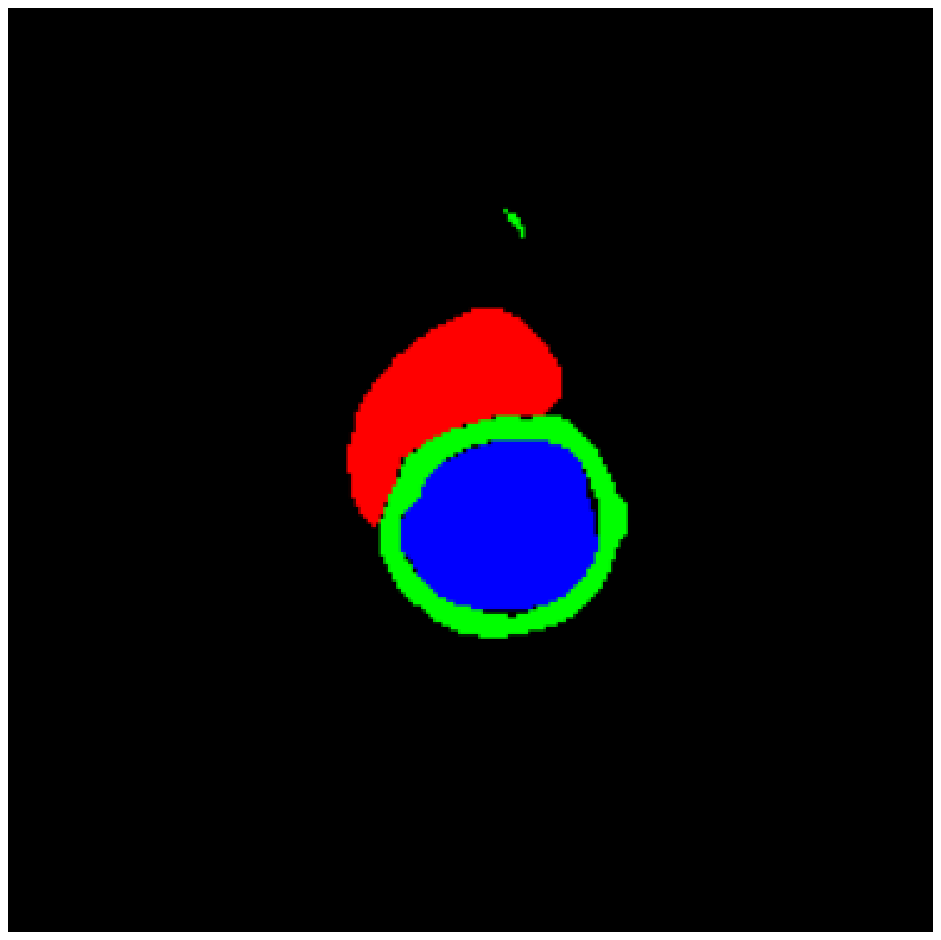}}
		\vspace{3pt}
		\centerline{\includegraphics[width=\textwidth]{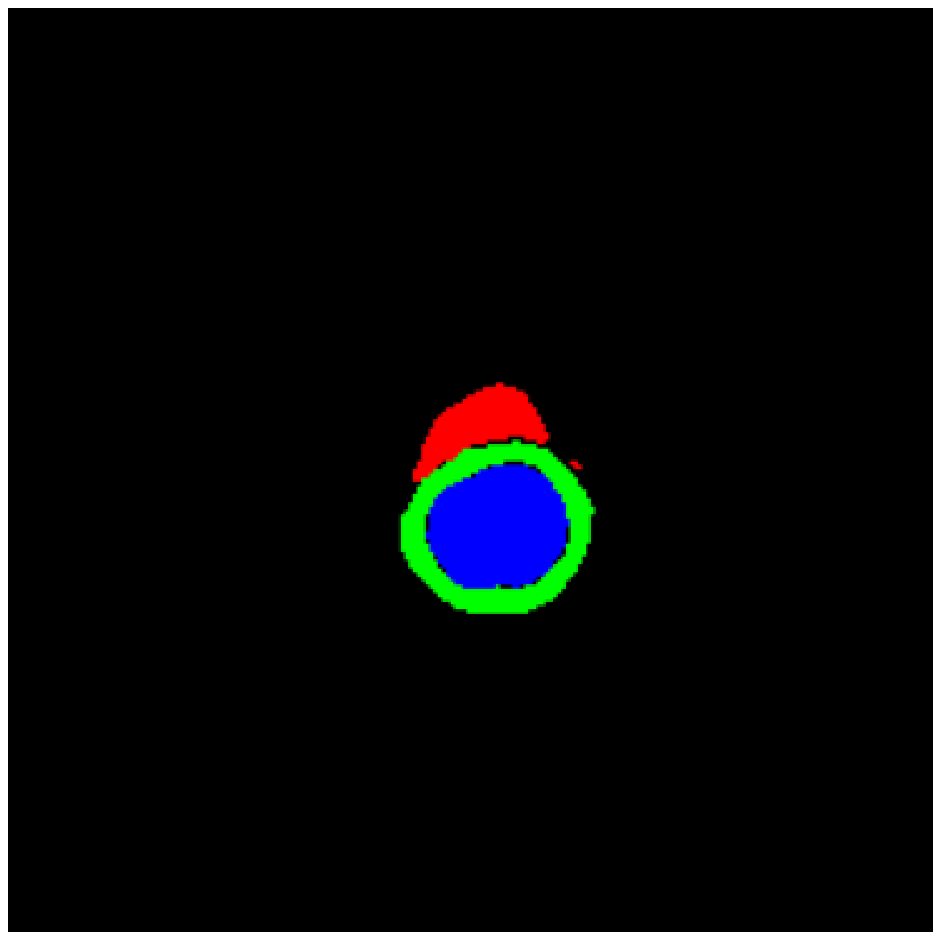}}
		\centerline{LGSA-U}
	\end{minipage}
	\begin{minipage}{0.15\linewidth}
		\vspace{3pt}
		\centerline{\includegraphics[width=\textwidth]{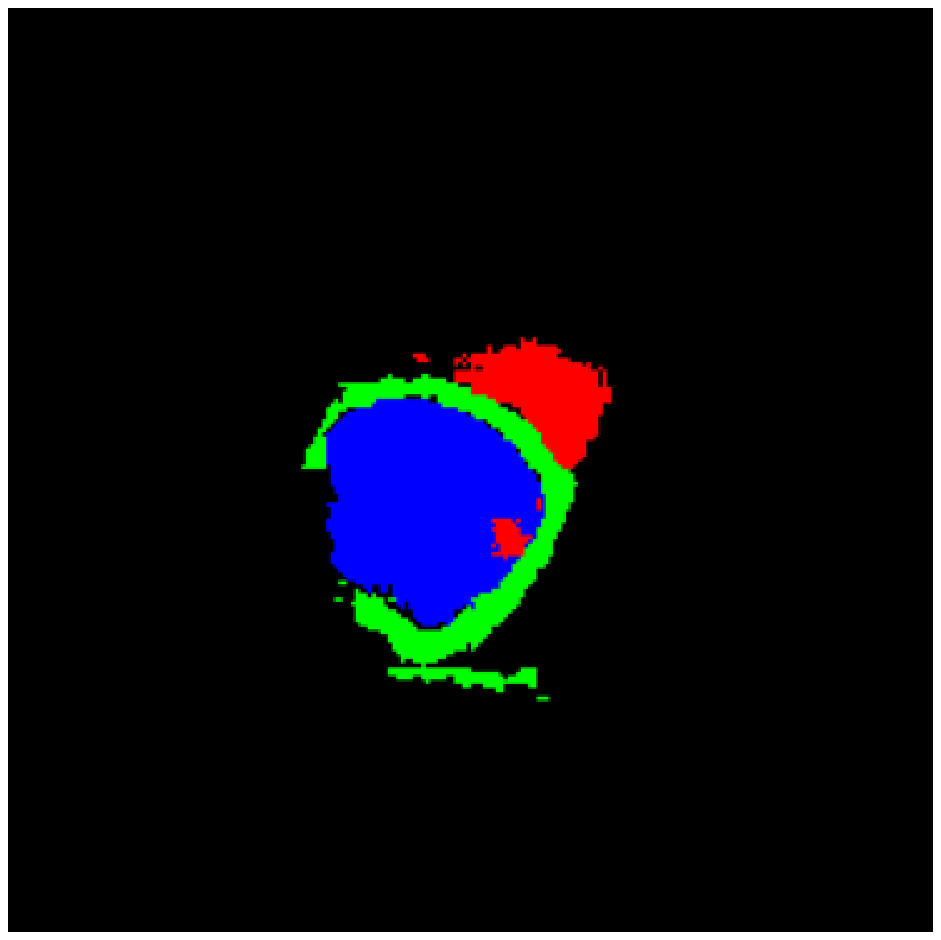}}
		\vspace{3pt}
		\centerline{\includegraphics[width=\textwidth]{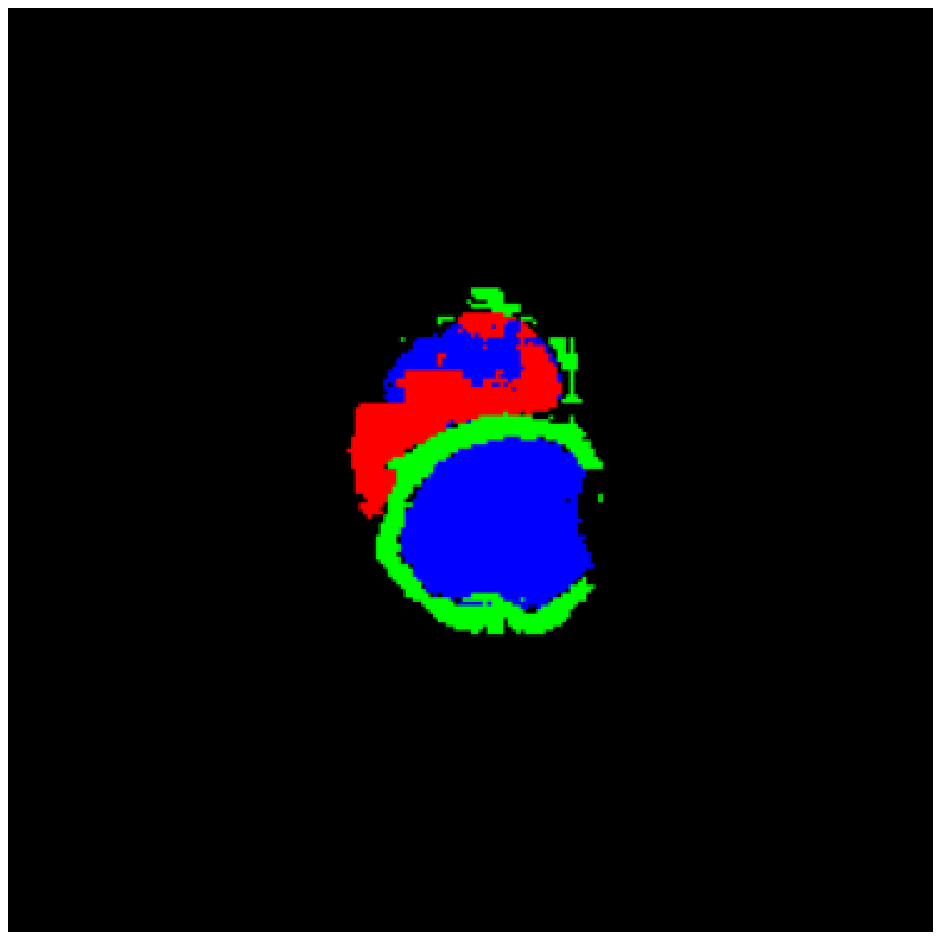}}
		\vspace{3pt}
		\centerline{\includegraphics[width=\textwidth]{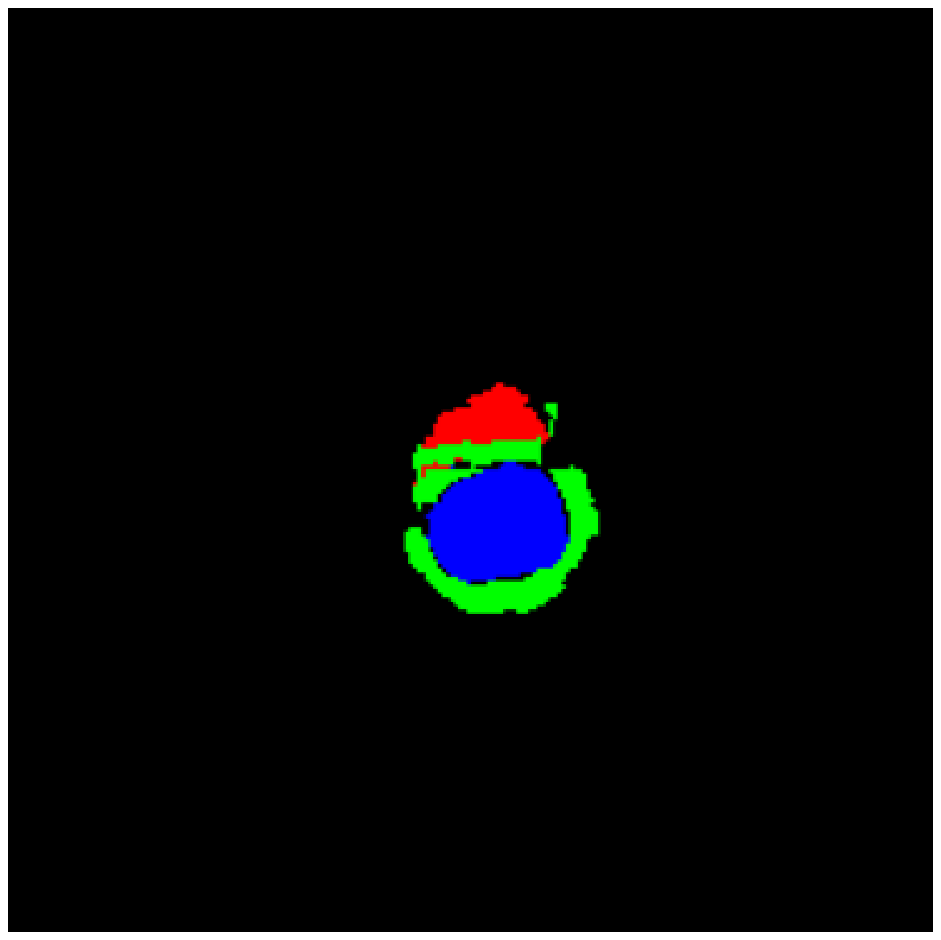}}
		\centerline{SwinUNet}
	\end{minipage}
	\begin{minipage}{0.15\linewidth}
		\vspace{3pt}
		\centerline{\includegraphics[width=\textwidth]{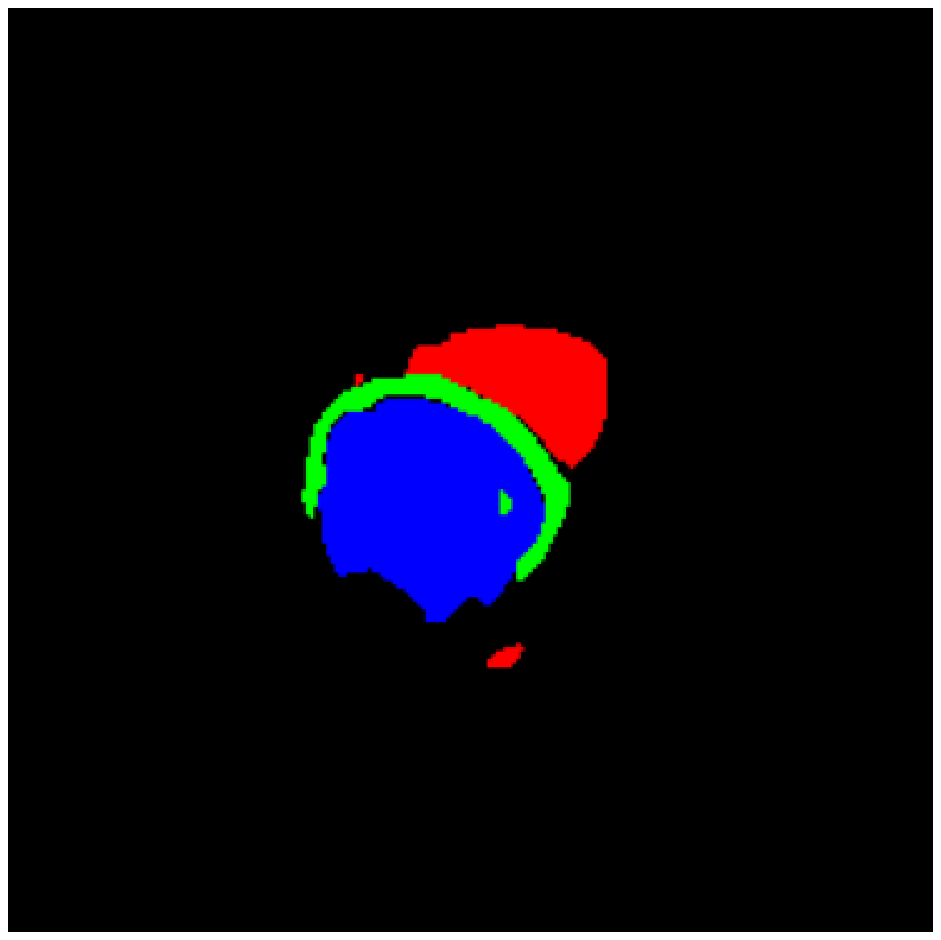}}
		\vspace{3pt}
		\centerline{\includegraphics[width=\textwidth]{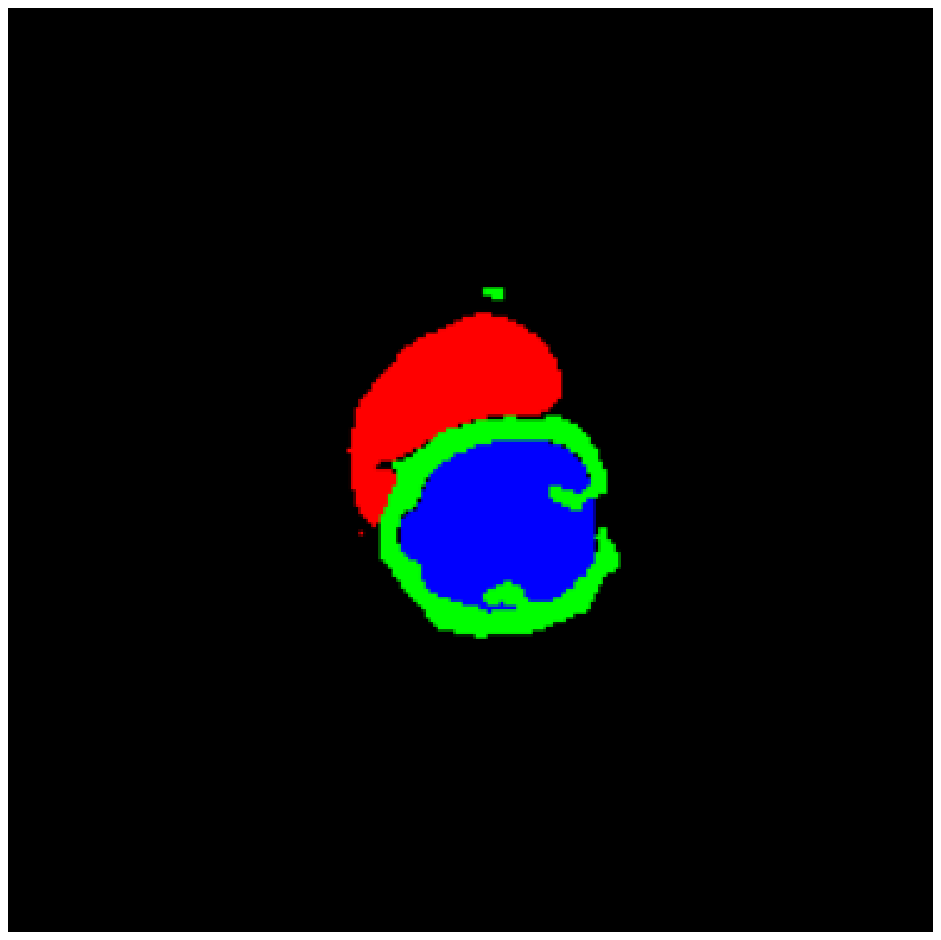}}
		\vspace{3pt}
		\centerline{\includegraphics[width=\textwidth]{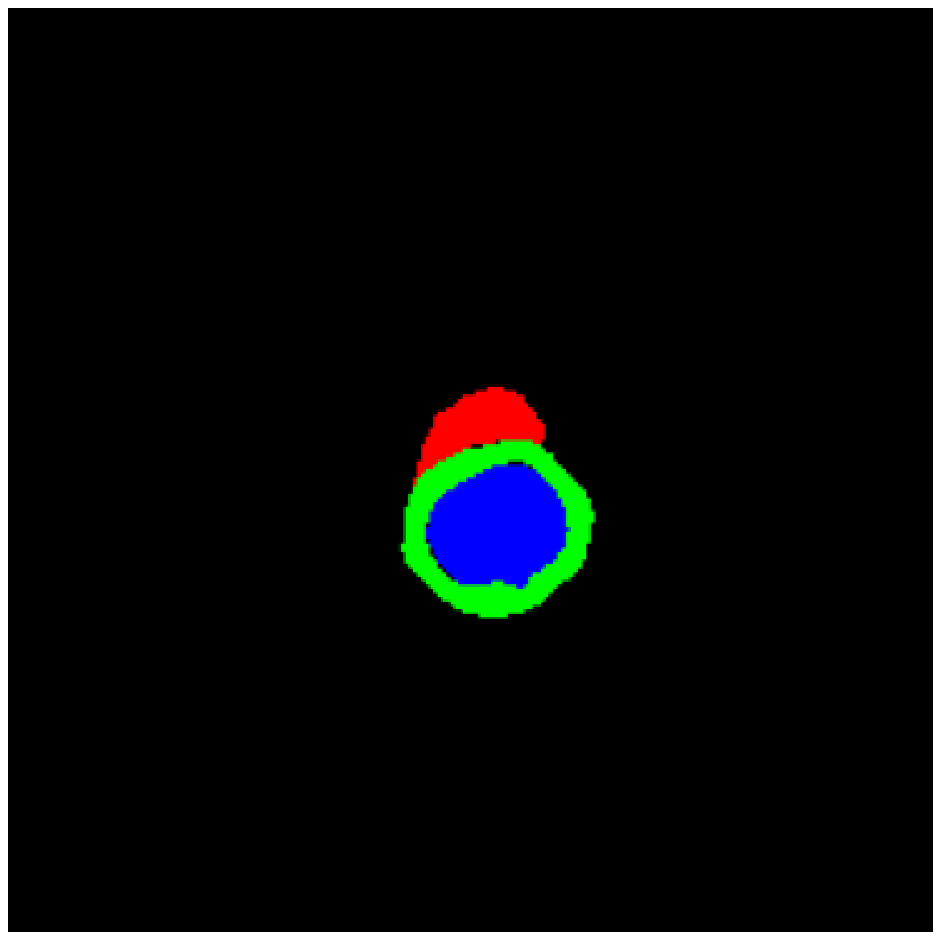}}
		\centerline{UNet}
	\end{minipage}
	\caption{Visualization of segmentation results using different methods in ACDC dataset. LGSA-S repesents LGSA-SwinUNet and LGSA-U repesents LGSA-UNet.
	}
\end{figure}

\begin{figure}[h]
	\centering
	\begin{minipage}{0.15\linewidth}
		\vspace{3pt}
		\centerline{\includegraphics[width=\textwidth]{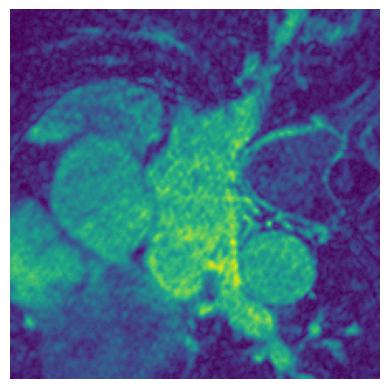}}
		\vspace{3pt}
		\centerline{\includegraphics[width=\textwidth]{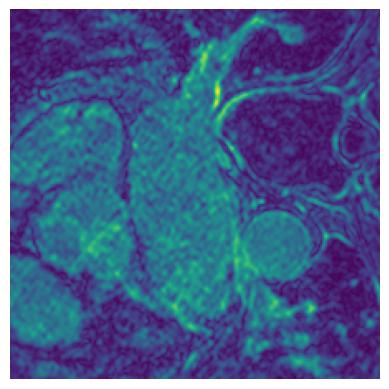}}
		\vspace{3pt}
		\centerline{\includegraphics[width=\textwidth]{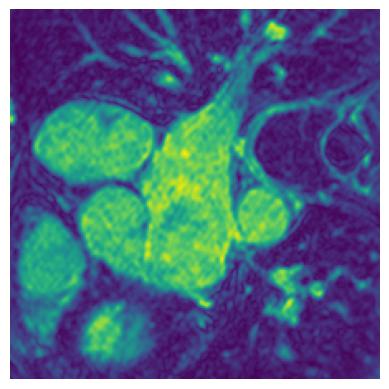}}
		\centerline{Image}
	\end{minipage}
	\begin{minipage}{0.15\linewidth}
		\vspace{3pt}
		\centerline{\includegraphics[width=\textwidth]{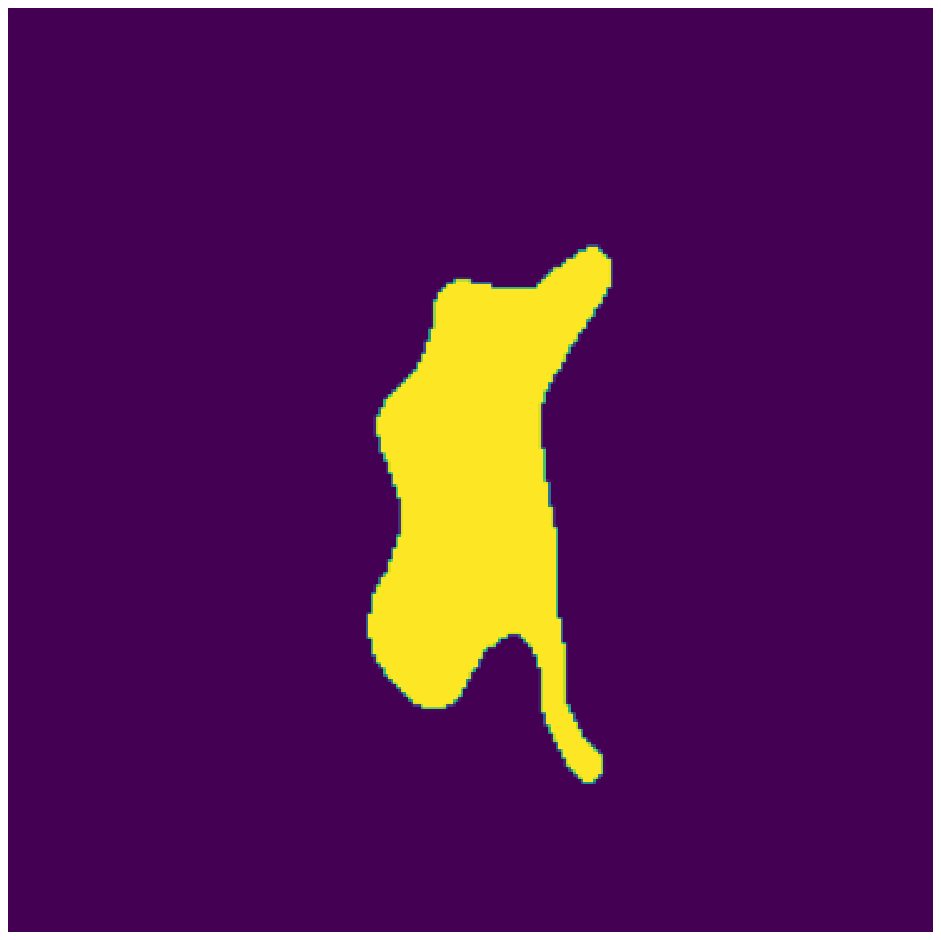}}
		\vspace{3pt}
		\centerline{\includegraphics[width=\textwidth]{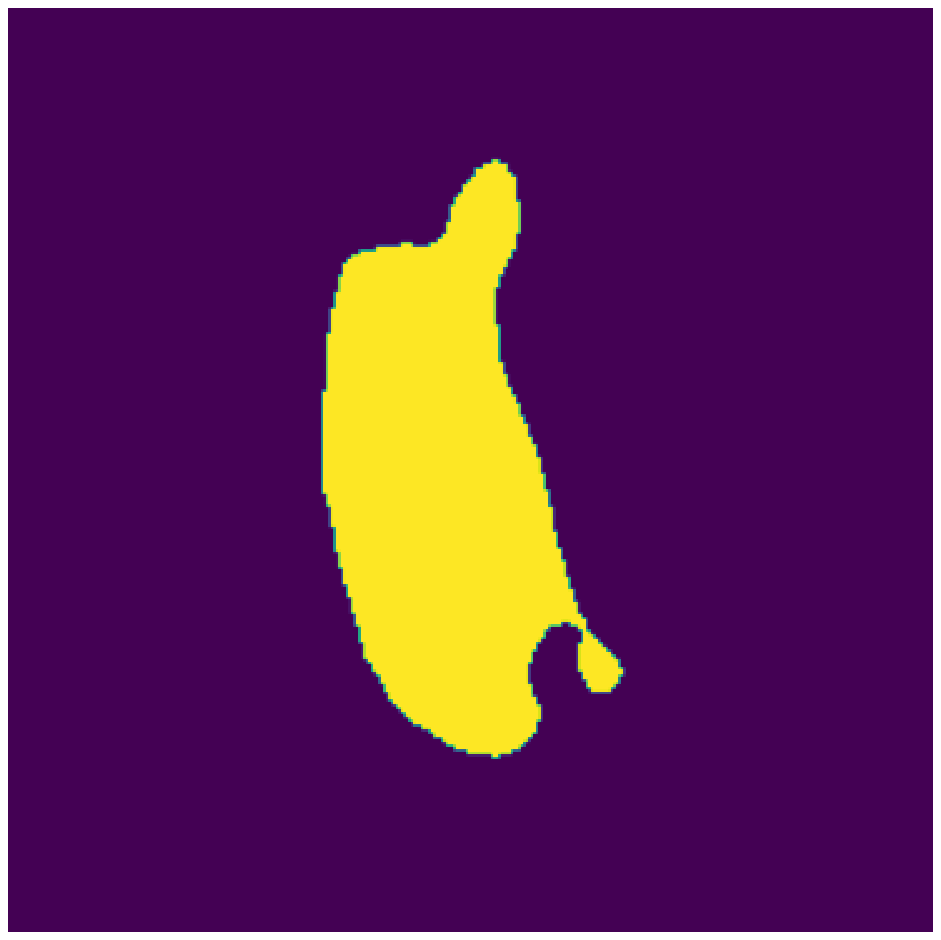}}
		\vspace{3pt}
		\centerline{\includegraphics[width=\textwidth]{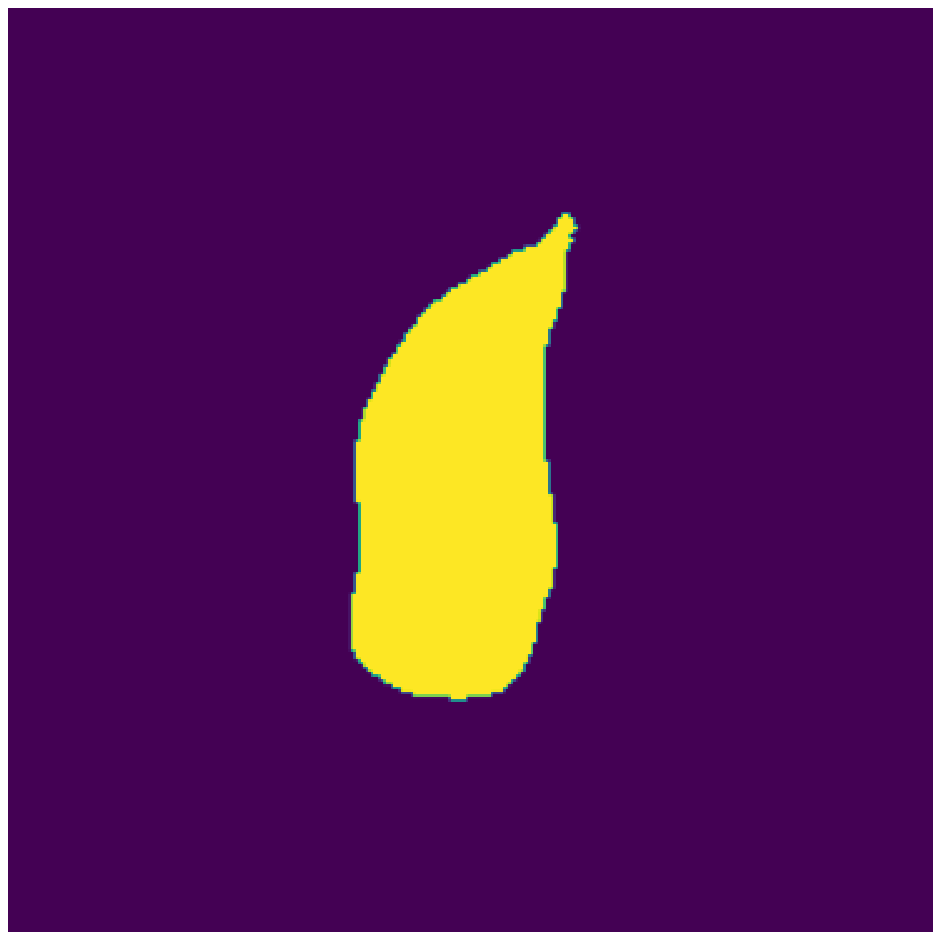}}
		\centerline{GroundTruth}
	\end{minipage}
	\begin{minipage}{0.15\linewidth}
		\vspace{3pt}
		\centerline{\includegraphics[width=\textwidth]{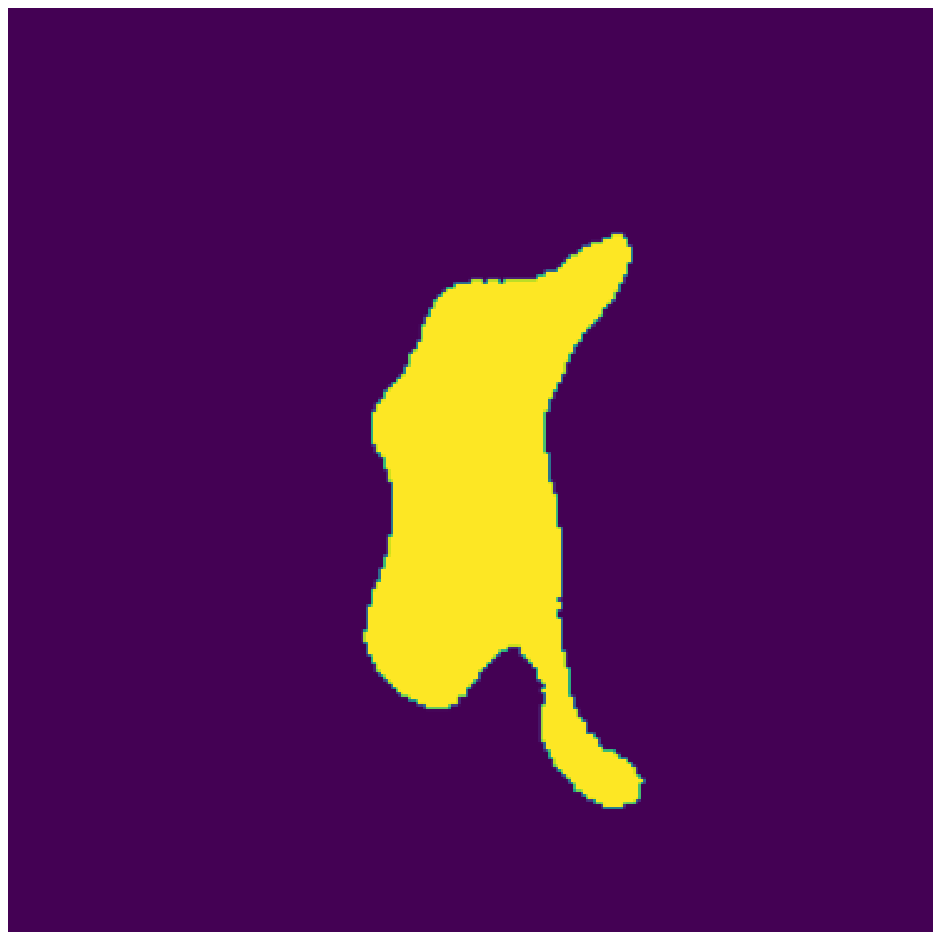}}
		\vspace{3pt}
		\centerline{\includegraphics[width=\textwidth]{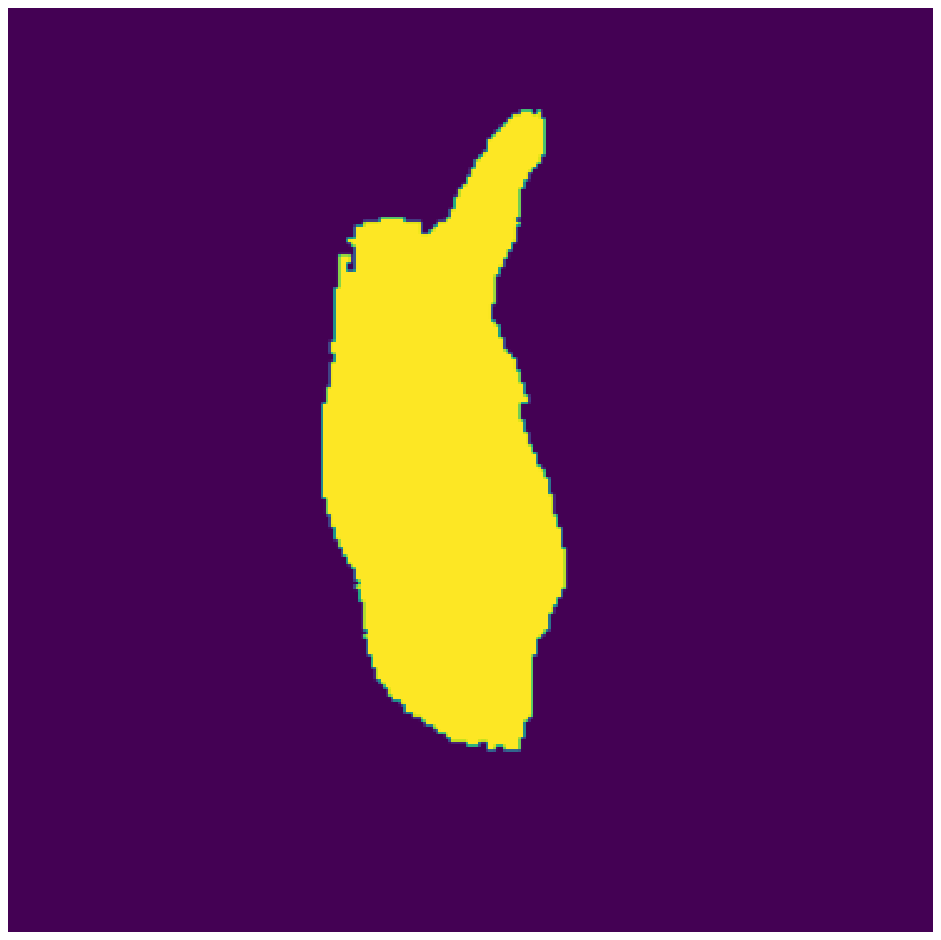}}
		\vspace{3pt}
		\centerline{\includegraphics[width=\textwidth]{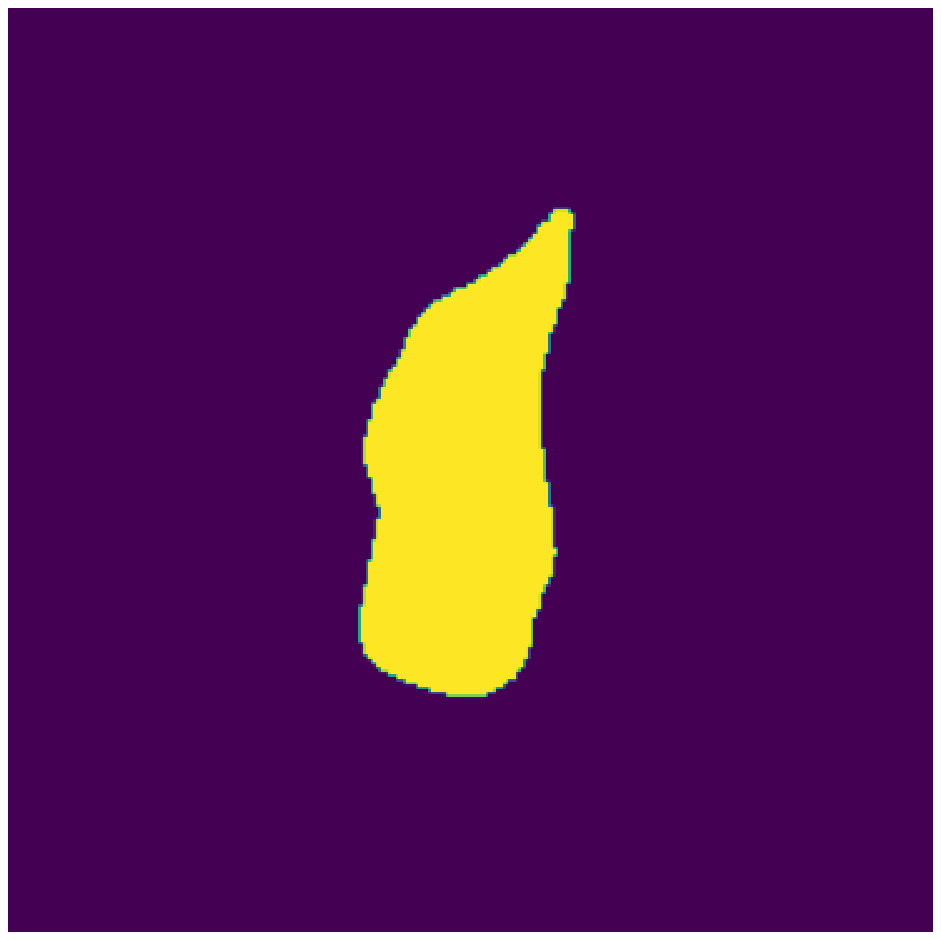}}
		\centerline{LGSA-S}
	\end{minipage}
	\begin{minipage}{0.15\linewidth}
		\vspace{3pt}
		\centerline{\includegraphics[width=\textwidth]{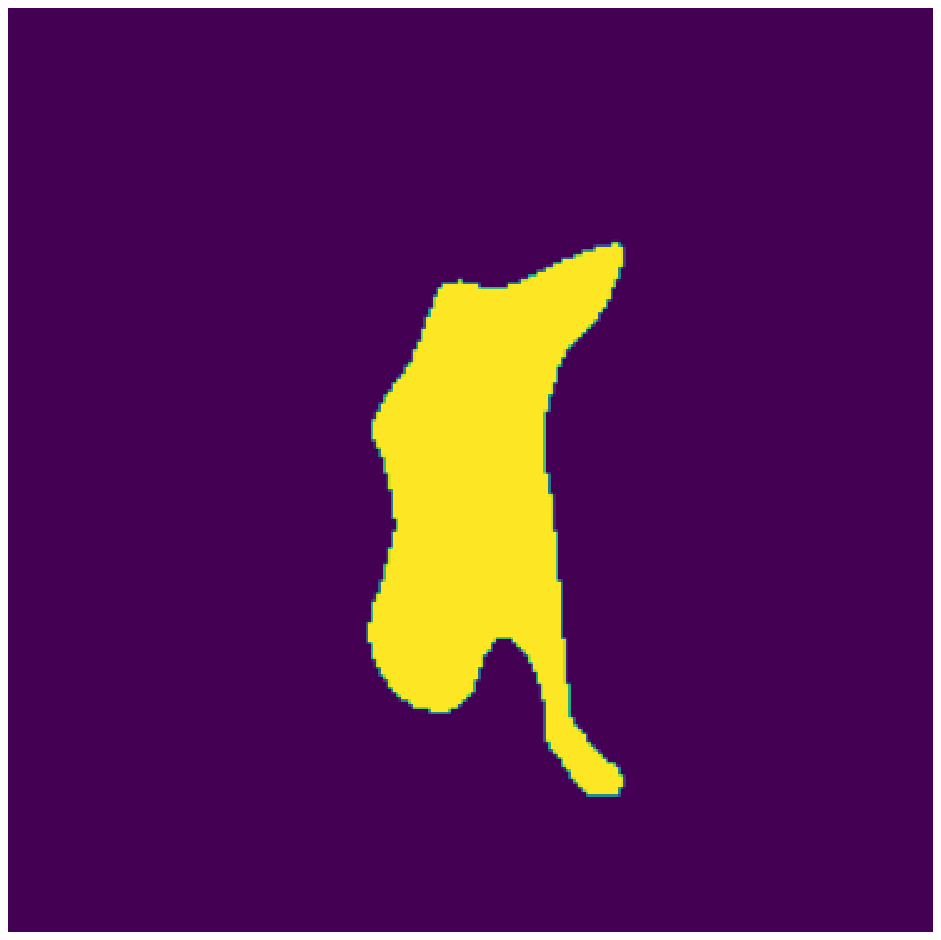}}
		\vspace{3pt}
		\centerline{\includegraphics[width=\textwidth]{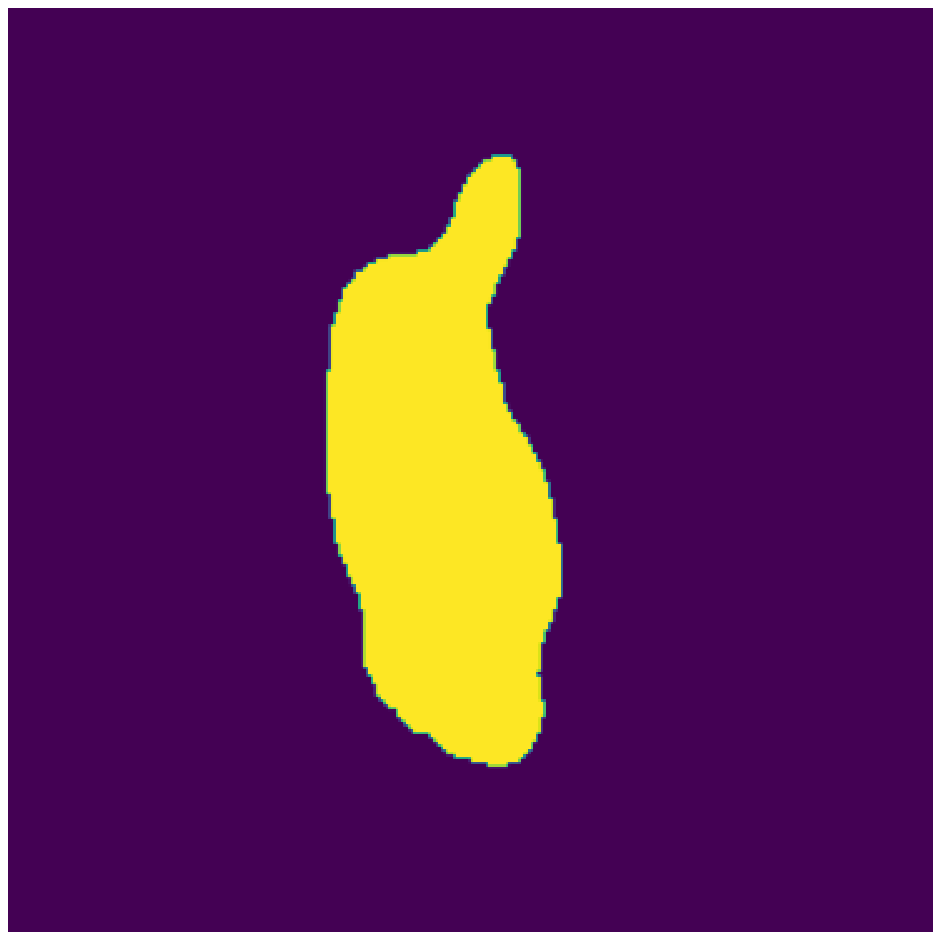}}
		\vspace{3pt}
		\centerline{\includegraphics[width=\textwidth]{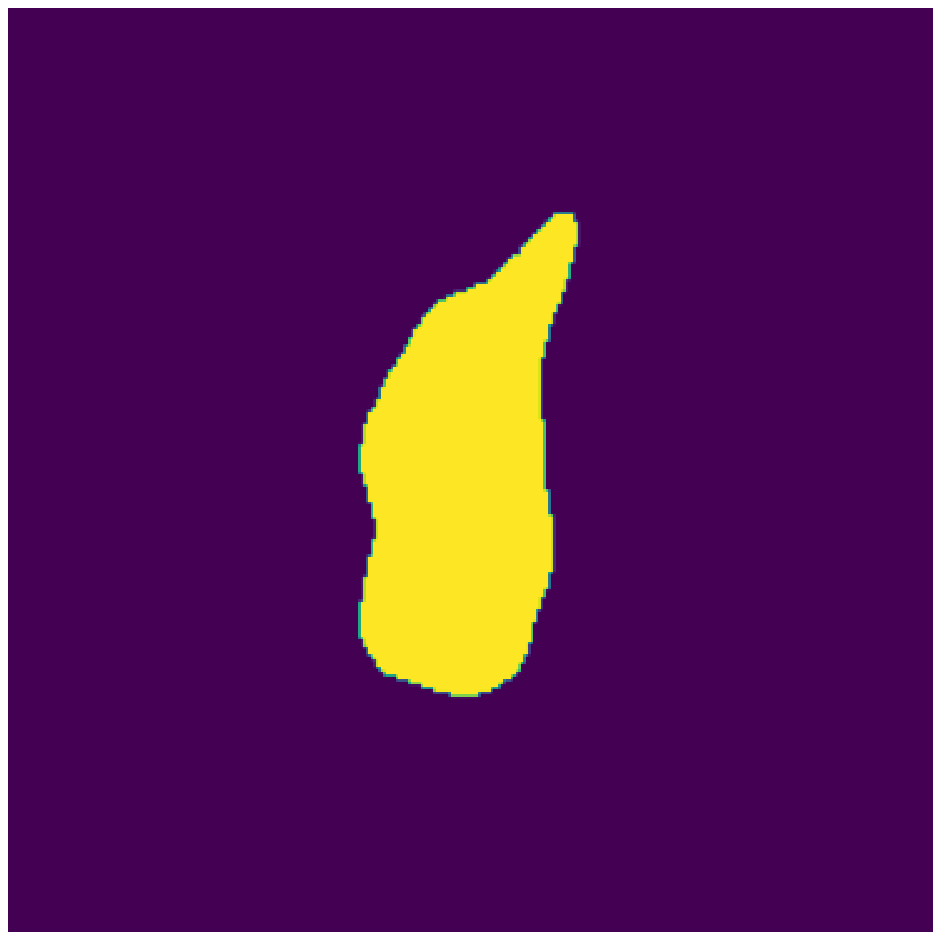}}
		\centerline{LGSA-U}
	\end{minipage}
	\begin{minipage}{0.15\linewidth}
		\vspace{3pt}
		\centerline{\includegraphics[width=\textwidth]{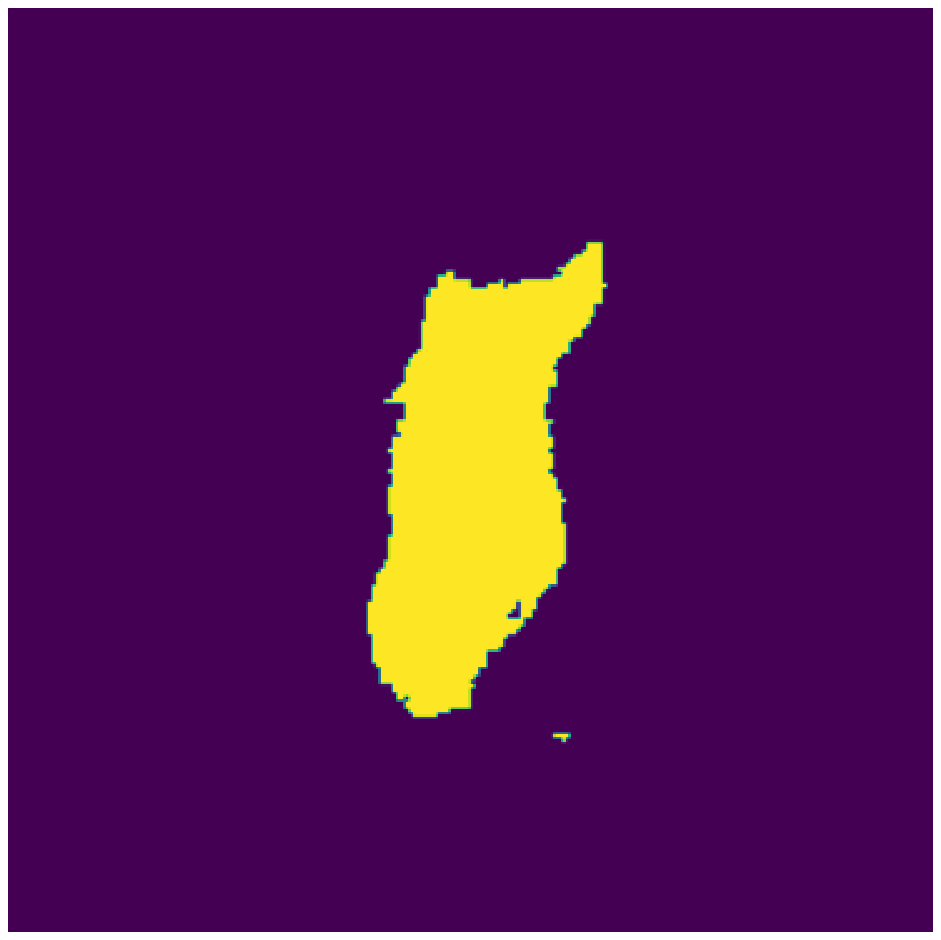}}
		\vspace{3pt}
		\centerline{\includegraphics[width=\textwidth]{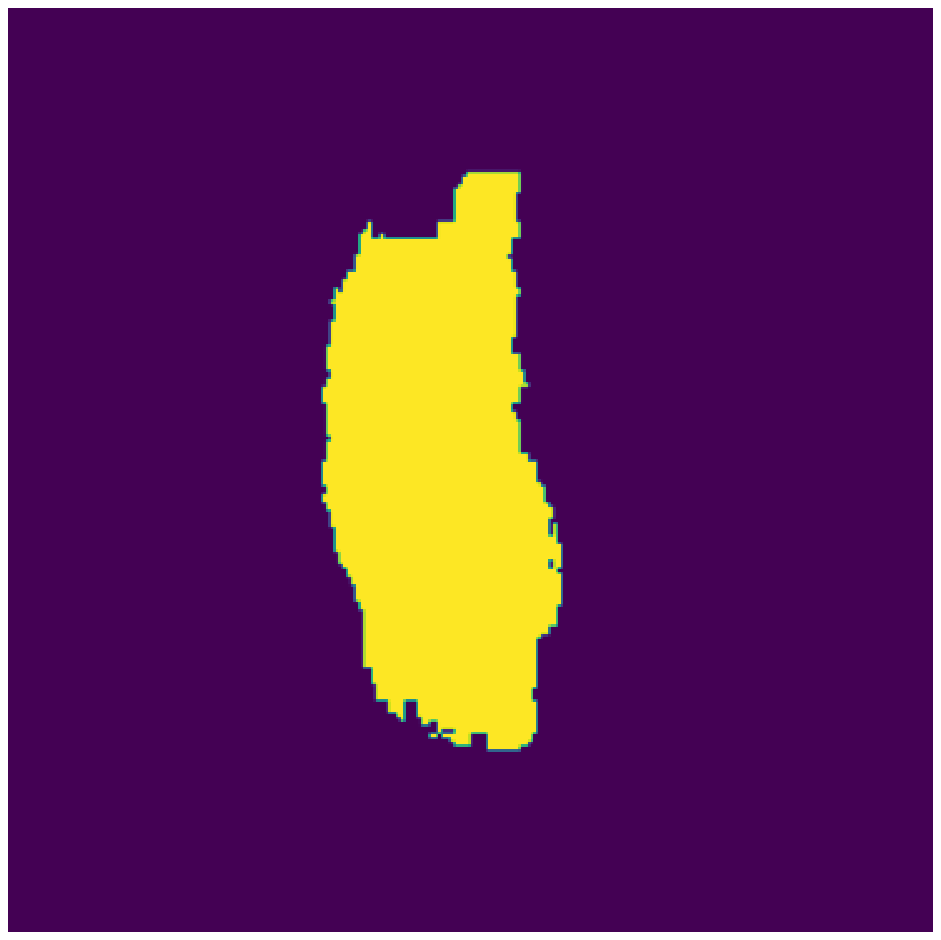}}
		\vspace{3pt}
		\centerline{\includegraphics[width=\textwidth]{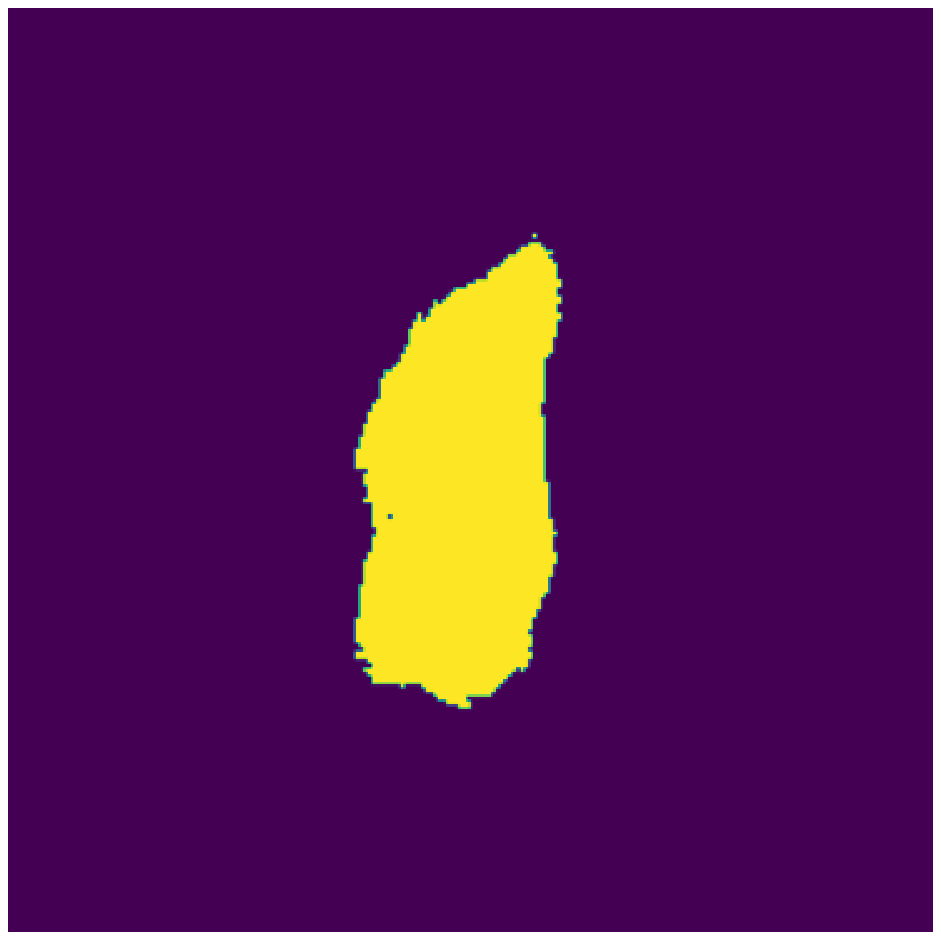}}
		\centerline{SwinUNet}
	\end{minipage}
	\begin{minipage}{0.15\linewidth}
		\vspace{3pt}
		\centerline{\includegraphics[width=\textwidth]{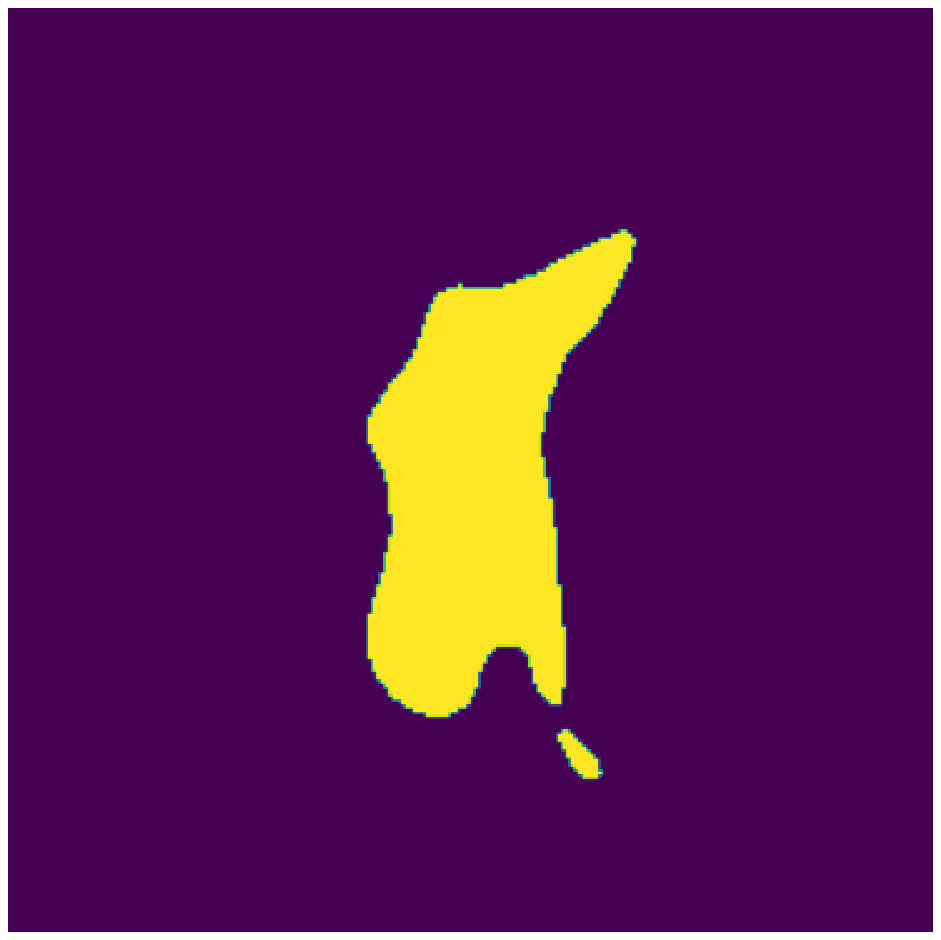}}
		\vspace{3pt}
		\centerline{\includegraphics[width=\textwidth]{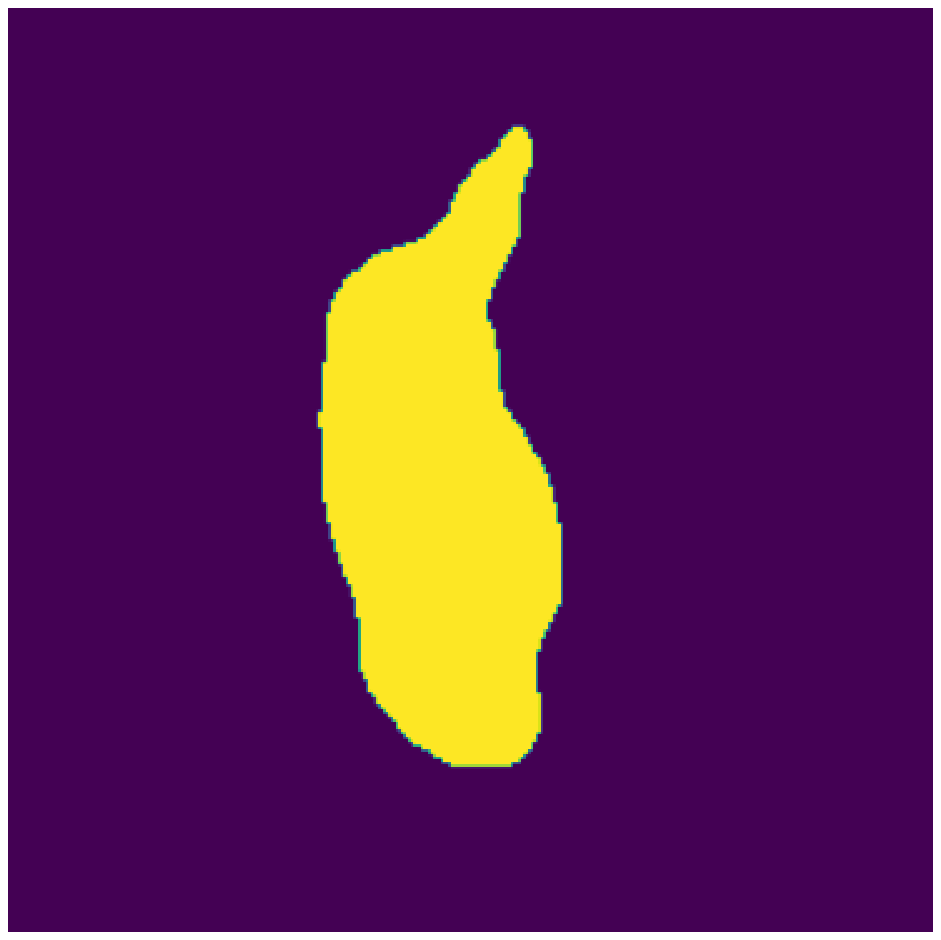}}
		\vspace{3pt}
		\centerline{\includegraphics[width=\textwidth]{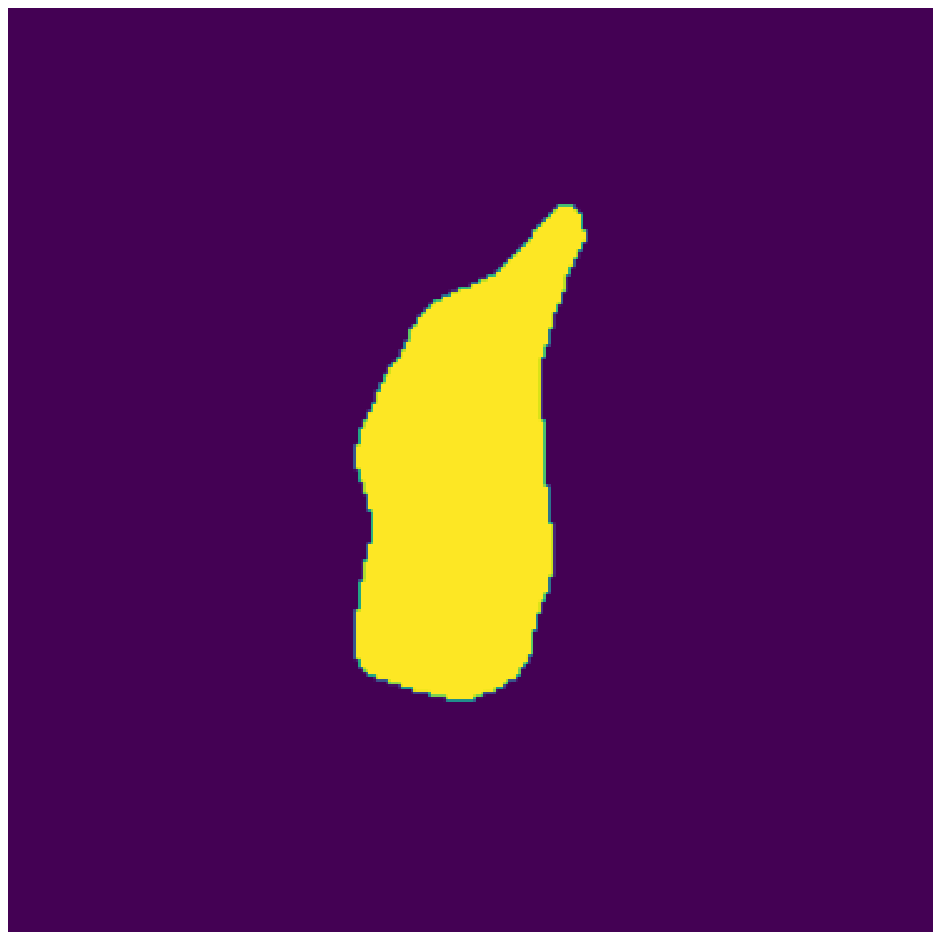}}
		\centerline{UNet}
	\end{minipage}
	\caption{Visualization of segmentation results using different methods in ASC dataset.LGSA-S repesents LGSA-SwinUNet and LGSA-U repesents LGSA-UNet.
	}
\end{figure}

\section{Conclusion and future work}
In this paper, we propose an atrium segmentation network based on location guidance and siamese adjustment, which takes consecutive three-layer slices as inputs.It uses location information in stage 1 to guide encoding features in stage 2, and conducts siamese interactions among the three-layer slices to take advantage of contextual information. We use a combination of serial supervision and siamese supervision to obtain the best optimization effect of this network. Experiments show that our method is suitable for classic 2D networks such as UNet, SwinUNet to achieve a significant performance improvement. In future work, we will further attempt to introduce edge detectors into segmentation tasks to improve the performance of segmentation.

\appendix
\section{Acknowledgements}
The author thanks the whole authors in the referred articles.
This work was supported in part by the Science and Technology 
Planning Project of Guangdong Science and Technology Department 
under Grant Guangdong Key Laboratory of Advanced IntelliSense 
Technology (2019B121203006). 

\clearpage
%
%
\bibliographystyle{splncs04}
\bibliography{egbib}
\end{document}